\documentclass[aps,prc,preprint,groupedaddress,nofootinbib]{revtex4-2}

\usepackage{graphicx}
\usepackage{amsmath,bm,ascmac,amssymb,calc}
\usepackage[dvipdfmx]{color}
\usepackage{CJK}

\begin{document}
\begin{CJK*}{UTF8}{}
\CJKfamily{min}


\title{Self-consistent single-nucleon potential at positive energy
  produced by semi-realistic interaction
  and its examination via nucleon-nucleus elastic scattering}


\author{H. Nakada (中田 仁)$^1$}
\email[E-mail:\,\,]{nakada@faculty.chiba-u.jp}
\author{K. Ishida (石田 佳香)$^2$}

\affiliation{$^1$ Department of Physics, Graduate School of Science,
 Chiba University,\\
Yayoi-cho 1-33, Inage, Chiba 263-8522, Japan}
\affiliation{$^2$ Department of Physics,
 Graduate School of Science and Engineering,
 Chiba University,\\
Yayoi-cho 1-33, Inage, Chiba 263-8522, Japan}


\date{\today}

\begin{abstract}
  Based on the variational principle,
  self-consistent single-particle (s.p.) potentials at positive energies
  are discussed,
  which correspond to the real part of the optical potential
  as the single folding potential (SFP).
  The nuclear-matter s.p. potential
  produced by the semi-realistic nucleonic interaction M3Y-P6,
  which has links to the bare nucleonic interaction,
  resembles those extracted from the empirical optical potential.
  Applying M3Y-P6 both to the self-consistent mean-field calculations
  for the target nucleus
  and to the SFP for the scattered nucleon,
  we find that the differential cross-sections
  of the nucleon-nucleus elastic scattering
  are reproduced almost comparably to the empirical potentials
  up to $80\,\mathrm{MeV}$ incident energy.
  The results demonstrate
  that the s.p. potential compatible with available experimental data
  can be derived from a single energy-independent effective interaction
  in this wide energy range.
\end{abstract}


\maketitle
\end{CJK*}



\section{Introduction\label{sec:intro}}

Many-body systems comprised of nucleons,
from atomic nuclei to objects in the universe,
are important ingredients of Nature.
While primarily governed by the strong interaction,
they often behave as quantum Fermi liquid~\cite{ref:BP91},
in which constituent nucleons move almost independently under a mean field (MF).
At the ground state (g.s.),
the nuclear MF contains correlation effects
as incorporated by the Brueckner theory~\cite{ref:FW71},
and is nowadays discussed in terms of the Kohn-Sham (KS) method
in the density functional theory~\cite{ref:PS91}.
In the KS method,
properties of the whole many-fermion system can be described
in terms of a collection of single-particle (s.p.) orbitals
under the self-consistent MF (SCMF)
constructed from effective interaction
(or energy density functional)~\cite{ref:KS65,ref:Nak23}.

Nuclear equation-of-state (EoS),
\textit{i.e.}, the energy of the nuclear matter
as a function of density and temperature ($T$)~\cite{CompOSE},
plays a vital role in supernovae and neutron stars.
Whereas the EoS at $T=0$ has been investigated relatively well
in connection to the experimental data
and the EoS at finite $T$ has often been developed
by extending it~\cite{ref:LS91,ref:Shen98,ref:SHT11},
the finite-$T$ EoS has not sufficiently been verified by experiments.
At finite $T$,
the constituent nucleons distribute over a wide energy range.
Therefore, it is desirable to handle the nucleonic states
without discontinuity with respect to energy.
The extension of the KS or the SCMF approaches
to finite $T$~\cite{ref:BLV84,ref:DLVM19},
in which the energy of the system is described by the s.p. states
obeying the Fermi-Dirac distribution function,
is a promising tool to obtain the EoS in connection to experimental data.
However, as the effective interactions have been examined
only via the nuclear structure data,
the reliability of these approaches has been limited
to low $T\,(\lesssim\,10\mathrm{MeV}$) so far.
In the supernovae,
$T$ may reach as high as $T_\mathrm{max}\sim 30\,\mathrm{MeV}$~\cite{ref:Sum21},
at which nucleons distribute over s.p. energies $\epsilon$
up to a few times $T_\mathrm{max}$.
Although some EoSs have been developed from the bare nucleonic interaction
through sophisticated theoretical methods~\cite{ref:Tog17,ref:BSVW21}
up to finite-$T$ cases~\cite{ref:BF99,ref:LLB19},
they are not easily compared with a variety of experimental data.
Since the energy distribution of nucleons is determined by the MF,
\textit{i.e.}, the s.p. potential produced by the nucleonic interaction,
it is a crucial step to examine whether the effective interaction produces
adequate s.p. potential at $\epsilon>0$,
as well as in the nuclear structure.

Nuclear MF at $\epsilon>0$ is connected
to the nucleon-nucleus ($N$-$A$) elastic scattering.
The $N$-$A$ elastic scattering is described by the optical potential
$\mathcal{U} = \mathcal{V} + i\,\mathcal{W}$~\cite{ref:Gle83,ref:Sat83},
where $\mathcal{V}$ and $\mathcal{W}$ are hermitian one-body operators,
often expressed by real functions of the position.
The imaginary part $\mathcal{W}$ carries effects of absorption,
\textit{i.e.}, loss of the flux out of the elastic channel.
Most conventionally, both $\mathcal{V}$ and $\mathcal{W}$ were adjusted
to the data.
A local function was assumed,
with the parameters dependent on the incident energy
and the mass number~\cite{ref:CH89,ref:KD03}.
The folding model has been developed to derive the optical potential
from the nucleonic effective interaction~\cite{ref:SL79},
which does not need parameters depending on the mass number.
There have been attempts to derive folding potentials
from the bare nucleonic interaction~\cite{ref:JLM76,ref:YNM86,ref:Amos00,
  ref:FSY08,ref:HKMW13,ref:VFG16,ref:WLH21}.

Under the thermal equilibrium,
there is no absorption
because of the detailed balance between inflow and outflow.
Only the real potential $\mathcal{V}$ is relevant,
involving correlation effects
like the s.p. potential in the KS theory.
In this respect, $\mathcal{V}$ is of particular interest,
which could be the s.p. potential at $\epsilon>0$
continuous with the MF potential in nuclear structure.
The Skyrme and the Gogny interactions,
which are effective interactions developed for nuclear SCMF calculations,
have been applied to the $N$-$A$ elastic scattering~\cite{ref:DG72,
  ref:BG79,ref:QJY81,ref:PKS10,ref:NDE11,ref:MO12,ref:HLP15,ref:Bla15,ref:LV21}.
However,
the good applicability of these phenomenological interactions
could be limited to low energy.
Whereas the imaginary potential has also been argued
within the many-body perturbation theory (MBPT)~\cite{ref:BG79,ref:QJY81,
  ref:MO12,ref:Bla15,ref:XGH14},
correlation effects already contained in the effective interaction
have yet to be subtracted.
If we can develop a reliable real potential
covering a wide energy range without counting on the MBPT,
it could be a significant step
toward a reasonable SCMF (or KS) approach at finite $T$.
It should be mentioned that a Brueckner-Hartree-Fock approach to EoS
combined with the $N$-$A$ scattering was reported in Ref.~\cite{ref:RSP13},
though not precisely examined by nuclear structure.

The Michigan-three-range-Yukawa (M3Y) interaction
was developed for $N$-$A$ inelastic scattering,
based on the $G$-matrix~\cite{ref:M3Y,ref:M3Y-P}.
By introducing density-dependent coefficients,
the M3Y interaction was extensively applied
to the folding potential~\cite{ref:DDM3Y,ref:BDM3Y}.
One of the authors (H.N.) evolved M3Y-type effective interactions
applicable to nuclear structure~\cite{ref:Nak03}.
In particular, the parameter-set M3Y-P6~\cite{ref:Nak13}
has been scrutinized in the SCMF approach~\cite{ref:Nak20},
and notable success has been found
in describing the nuclear shell structure~\cite{ref:Nak20,ref:NS14},
establishing reliability for s.p. potential at $\epsilon<0$.
Moreover,
M3Y-P6 is compatible with the EoS parameters at $T=0$ extracted by experiments
and reproduces a microscopic neutron-matter EoS~\cite{ref:Nak13,ref:Nak20}.
It has also been pointed out
that the M3Y-type interactions are almost free from
unphysical instabilities in excitations of the nuclear matter~\cite{ref:DPN21},
unlike many other MF interactions.
A SCMF approach with M3Y-P6 has been extended to finite $T$
to investigate the liquid-gas phase transition
occurring at $T\approx 10\,\mathrm{MeV}$ in Ref.~\cite{ref:DLVM19}.
It is interesting to examine this effective interaction
for the $N$-$A$ scattering.

\section{Single folding potential
  and self-consistent mean field\label{sec:SFP}}

Within the SCMF scheme,
the total energy $E$
is represented by
\begin{equation}
  E = \sum_\alpha \langle\alpha|\frac{\mathbf{p}^2}{2M}|\alpha\rangle\,n_\alpha
  + \frac{1}{2}\sum_{\alpha \beta}
  \langle\alpha\beta|\hat{v}|\alpha\beta\rangle\,n_\alpha\,n_\beta\,.
\label{eq:tot-energy}\end{equation}
The indices $\alpha$ and $\beta$ denote s.p. states,
which will be commonly used for labeling nucleons without confusion,
$n_\alpha$ is the occupation probability on $\alpha$,
and $\hat{v}$ is the two-body interaction,
whose strengths may depend on the density.
The s.p. Hamiltonian $h$ is derived as
\begin{equation}
  h = \sum_\alpha \bigg[\Big(\frac{1}{n_\alpha}\,
  \frac{\delta E}{\delta\langle\alpha|}\Big)
  \bigg\vert_{n^{(0)}} \langle\alpha|\bigg]
  = \frac{\mathbf{p}^2}{2M} + U\,,
  \label{eq:sp-hamil}\end{equation}
from which the s.p. state $|\alpha\rangle$ and its energy $\epsilon_\alpha$
are obtained via $h|\alpha\rangle = \epsilon_\alpha|\alpha\rangle$.
We have defined $(\delta/\delta\langle\alpha|)
\langle\alpha'|\hat{\mathcal{O}}|\beta\rangle
= \delta_{\alpha\alpha'}\,\hat{\mathcal{O}}|\beta\rangle$
for a matrix element of a one-body operator $\hat{\mathcal{O}}$,
and analogously for two-body matrix elements.
The expression $\vert_{n^{(0)}}$ means
substituting appropriate values $n_\beta^{(0)}$ for $n_\beta$.
The second term on the right-hand side (rhs) of Eq.~\eqref{eq:tot-energy}
leads to the s.p. potential $U$,
which is non-local in general.
For spherical nuclei,
it is appropriate to take $\alpha=(\nu_r \ell jm\tau)$,
where $\tau\,(=p,n)$ is the particle type,
$(\ell jm)$ are the angular-momentum quantum numbers,
and $\nu_r$ distinguishes radial wave-functions.
For homogeneous nuclear matter,
we take $\alpha=(\mathbf{k}\sigma\tau)$,
with the momentum $\mathbf{k}$
and the $z$-component of the nucleon spin $\sigma$.


Suppose that $\hat{v}$ can be expressed as
\begin{equation}
  \hat{v} = \sum_i C_i[\rho]\cdot\hat{w}_i\,,
\label{eq:v_DD}\end{equation}
where $\hat{w}_i$ is a two-body operator
with strength $C_i$ that may depend on
$\rho(\mathbf{r})$,
as the interaction of Eq.~\eqref{eq:effint} in Appendix~\ref{app:hamil}.
Then, the s.p. potential $U$ in Eq.~\eqref{eq:sp-hamil} becomes
\begin{equation}\begin{split}
    U|\alpha\rangle =& \sum_i \sum_\beta \big\langle\!\!\ast\!\beta\big|
    C_i[\rho^{(0)}(\mathbf{R}_{\alpha\beta})]\cdot\hat{w}_i
    \big|\alpha\beta\big\rangle\,n_\beta^{(0)} \\
    &+ \frac{1}{2}\,|\alpha\rangle\,\sum_i \sum_{\alpha'\beta}
    C'_i[\rho^{(0)}(\mathbf{r}_\alpha)]\,\big\langle\alpha'\beta\big|
    \delta(\mathbf{r}_\alpha-\mathbf{R}_{\alpha'\beta})\cdot\hat{w}_i
    \big|\alpha'\beta\big\rangle\,n_{\alpha'}^{(0)}\,n_\beta^{(0)}\,.
\end{split}\label{eq:sp-pot}\end{equation}
We have assumed that $\rho$ depends
on $\mathbf{R}_{\alpha\beta}:=(\mathbf{r}_\alpha+\mathbf{r}_\beta)/2$
when it acts on two nucleons $\alpha$ and $\beta$.
The expression $\langle\ast\beta|$ means
that it is the result of the variation,
$\langle\ast\beta|:=(\delta/\delta\langle\alpha|)\langle\alpha\beta|$,
and $\rho^{(0)}$ is the density obtained by $n^{(0)}$.
The second term on the rhs that includes $C'_i=dC_i/d\rho$
is the rearrangement potential,
for which we have inserted $\delta\rho(\mathbf{r})/\delta\langle\alpha|
=\delta(\mathbf{r}_\alpha-\mathbf{r})\,|\alpha\rangle$.

In the homogeneous nuclear matter,
Eq.~\eqref{eq:sp-pot} arrives at
\begin{equation}\begin{split}
  & \langle\mathbf{k}\sigma\tau|U|\mathbf{k}\sigma\tau\rangle\\
  &\quad = \sum_i C_i[\rho]\,\frac{\Omega}{(2\pi)^3}\sum_{\sigma'\tau'}
 \int_{k'\leq k_{\mathrm{F}\tau'}}d^3k'\,
 \langle\mathbf{k}\sigma\tau\,\mathbf{k}'\sigma'\tau'|\hat{w}_i
 |\mathbf{k}\sigma\tau\,\mathbf{k}'\sigma'\tau'\rangle \\
 &\qquad + \sum_i C'_i[\rho]\,\frac{\Omega}{2(2\pi)^6}
 \sum_{\sigma\tau\sigma'\tau'}
 \int_{k\leq k_{\mathrm{F}\tau},k'\leq k_{\mathrm{F}\tau'}}d^3k\,d^3k'\,
 \langle\mathbf{k}\sigma\tau\,\mathbf{k}'\sigma'\tau'|\hat{w}_i
 |\mathbf{k}\sigma\tau\,\mathbf{k}'\sigma'\tau'\rangle\,.
\end{split}\label{eq:U-matter}\end{equation}
Here $\Omega$ is the volume of the system,
and $k_{\mathrm{F}\tau}$ $(\tau=p,n)$ denotes the Fermi momentum
that is related to the density,
$\rho_\tau=k_{\mathrm{F}\tau}^3/(3\pi^2)$ and $\rho=\sum_\tau \rho_\tau$.
Analytic formulae for the integration in Eq.~\eqref{eq:U-matter} are given
in Ref.~\cite{ref:Nak03}.
The potential $\langle\mathbf{k}\sigma\tau|U|\mathbf{k}\sigma\tau\rangle$
depends on $\rho$ and the asymmetry parameter $\eta_t$,
where $\eta_t:=\sum_\tau \tau \rho_\tau\big/\rho$
with $\tau=\pm 1$ in the summation,
as well as on $k=|\mathbf{k}|$ and $\tau$.

Let us consider the $N$-$A$ elastic scattering,
to which the above formulae are applicable.
The incident nucleon is denoted by $N$ with the energy $\epsilon_N$
(the subscript $N$ will occasionally be replaced by $p$ or $n$
in Sec.~\ref{sec:DCS}),
and the target nucleus by its mass number $A$,
whose g.s. energy and density are expressed as $E_A$ and $\rho_A^{(0)}$.
The occupation probabilities are $n_N^{(0)}=1$,
$n_\alpha^{(0)}=1$ for $\alpha$ belonging to the occupied states of $A$,
and $n_\alpha^{(0)}=0$ for all the other s.p. states.
Whereas the increment of the density due to the scattered nucleon
is infinitesimal at each position,
its variation with respect to $\langle N|$
is not negligible~\cite{ref:NS06}.
Therefore, the s.p. potential of Eq.~\eqref{eq:sp-pot} for $N$ is given by
\begin{equation}\begin{split}
    U|N\rangle =& \sum_i \sum_{\beta=1}^A
    \big\langle\!\!\ast\!\beta\big|C_i[\rho_A^{(0)}(\mathbf{R}_{N\beta})]\cdot
    \hat{w}_i\big|N\beta\big\rangle\,n_\beta^{(0)} \\
    &+ \frac{1}{2}\,|N\rangle\,\sum_i \sum_{\alpha,\beta=1}^A
    C'_i[\rho_A^{(0)}(\mathbf{r}_N)]\,
    \big\langle\alpha\beta\big|\delta(\mathbf{r}_N-\mathbf{R}_{\alpha\beta})\cdot
    \hat{w}_i\big|\alpha\beta\big\rangle\,n_\alpha^{(0)}\,n_\beta^{(0)}\,.
\end{split}\label{eq:sp-pot_N}\end{equation}
This $U$ corresponds to the single folding potential (SFP),
which generally has non-locality, owing to the exchange term.
The incident energy is $\epsilon_N = E-E_A
= \langle N|h|N\rangle/\langle N|N\rangle$.
In addition to the first term on the rhs of Eq.~\eqref{eq:sp-pot_N},
which is the conventional SFP,
the rearrangement potential appears in the second term~\cite{ref:HM72,ref:NS06,
  ref:Koh18,ref:LKP20}.
When three-body interaction acts on the system,
its effects are treated analogously.
If correlation effects are embodied in the effective interaction
as in the KS theory~\cite{ref:Nak23},
the above $U$ can be identified with the real part of the optical potential
$\mathcal{V}$.
Relativistic effects may partly be incorporated
into the effective interaction~\cite{ref:YNM86}, as well.
We denote $\mathcal{V}$ by $\mathcal{V}^\mathrm{SFP}$
when calculated via Eq.~\eqref{eq:sp-pot_N}.
The present derivation elucidates
that the SFP is a self-consistent s.p. potential at positive energies,
in complete analogy to the SCMF potential.
It deserves noting that $U$ in Eq.~\eqref{eq:sp-pot_N}
does not depend on $\epsilon_N$
when the non-locality is explicitly taken into account,
as will be confirmed from the formulae in Appendix~\ref{app:SFP}.

\section{Single-particle potential in nuclear matter\label{sec:matter}}

\begin{figure}
  \includegraphics[scale=0.5]{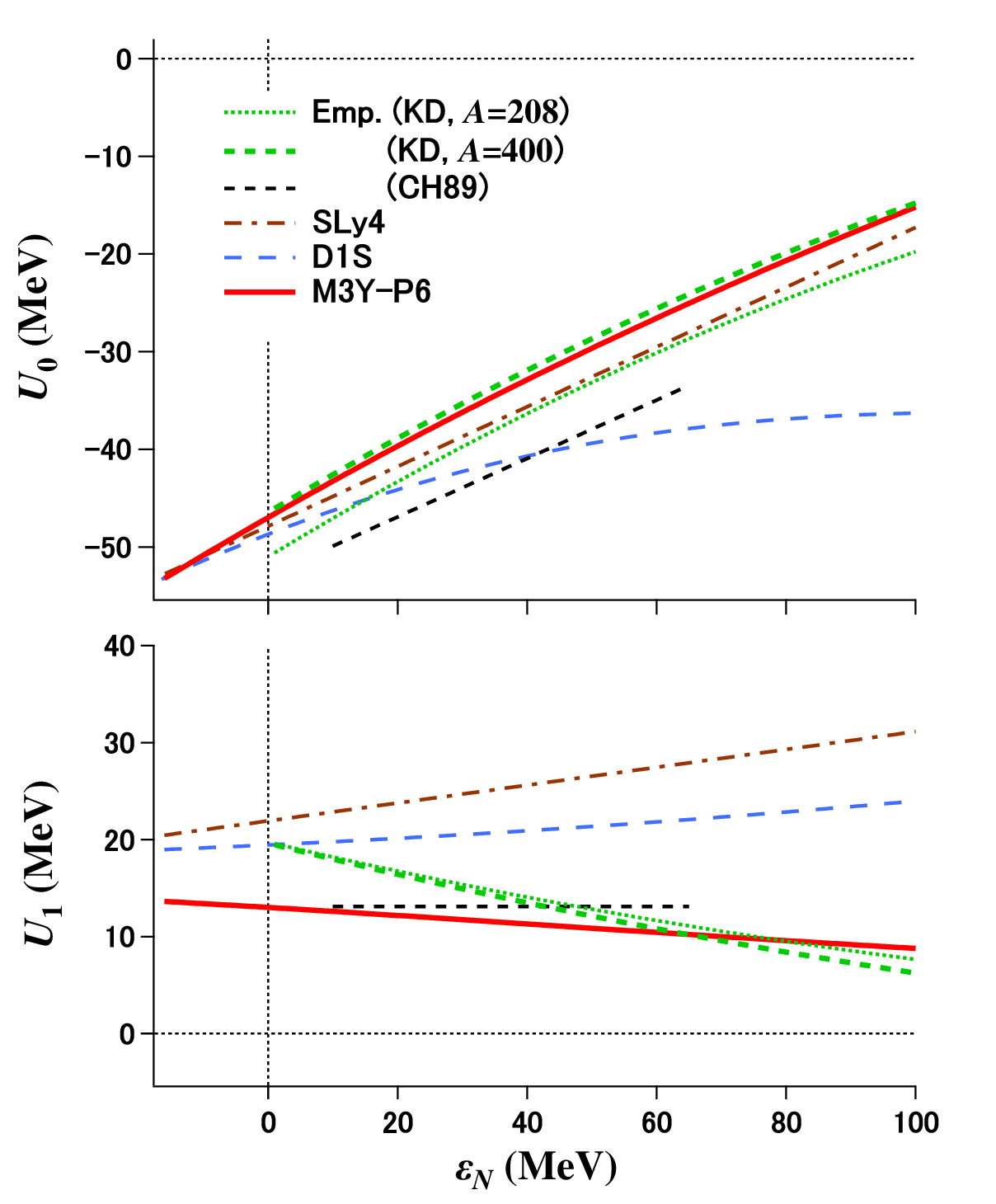}\vspace*{3mm}
  \caption{Energy dependence of $U_0$ and $U_1$ for homogeneous nuclear matter
    [see Eq.~\protect\eqref{eq:U-matter2}] at $\rho=0.16\,\mathrm{fm}^{-3}$.
    The s.p. potentials given by the SCMF interactions,
    Skyrme-SLy4 (brown dot-dashed lines), Gogny-D1S (blue long-dashed lines)
    and M3Y-P6 (red solid lines),
    are compared with the empirical potentials,
    KD evaluated at $A=208$ (green dotted lines),
    at $A=400$ (green short-dashed lines)
    and CH89 (black short-dashed lines).
\label{fig:U_NM}}
\end{figure}

In the homogeneous nuclear matter,
$\epsilon_N = k^2/(2M)
+ \langle\mathbf{k}\sigma\tau|U|\mathbf{k}\sigma\tau\rangle$
with the potential of Eq.~\eqref{eq:U-matter}.
The non-locality in $U$ can be absorbed in the momentum-dependence,
which is further translated into the $\epsilon_N$-dependence
without approximation,
because the momentum $\mathbf{k}$ is a good quantum number.
If the nuclear-matter energy is a quadratic function of $\eta_t$
to good approximation~\cite{ref:TN17},
the s.p. potential is represented as
\begin{equation}
  \langle\mathbf{k}\sigma\tau|U|\mathbf{k}\sigma\tau\rangle
  \approx U_0(\epsilon_N;\rho) + \tau\,U_1(\epsilon_N;\rho)\,\eta_t\,,
\label{eq:U-matter2}\end{equation}
corresponding to the Lane form~\cite{ref:Sat83}.
The s.p. potential at the saturation density $\rho_0$ can be compared
to the empirical local potential $\mathcal{V}^\mathrm{emp}$
at the center of heavy nuclei,
ideally the $A\to\infty$ limit,
\begin{equation}
  \lim_{A\to\infty} \mathcal{V}^\mathrm{emp}(\mathbf{r}=0)
  \approx U_0^\mathrm{emp}(\epsilon_N;\rho_0)
  - \tau\,U_1^\mathrm{emp}(\epsilon_N;\rho_0)\,\frac{N-Z}{A}\,.
  \quad\left(\tau=\left\{\begin{array}{ll}+1&\mbox{for $p$}\\
  -1&\mbox{for $n$}\end{array}\right.\right)
\end{equation}
In Fig.~\ref{fig:U_NM},
$U_0$ and $U_1$ at $\rho=0.16\,\mathrm{fm}^{-3}$
are shown as a function of $\epsilon_N$.
Several effective interactions that successfully describe nuclear structure
are applied:
Skyrme-SLy4~\cite{ref:SLy}, Gogny-D1S~\cite{ref:D1S}
and M3Y-P6~\cite{ref:Nak13}.
$U_0^\mathrm{emp}$ and $U_1^\mathrm{emp}$ are displayed for comparison,
for which we take those of Refs.~\cite{ref:CH89} (CH89)
and \cite{ref:KD03} (KD).
The CH89 potential was fitted to the data
in $10\leq \epsilon_N\leq 65\,\mathrm{MeV}$.
The KD potential, applicable in $0.001\leq \epsilon_N\leq 200\,\mathrm{MeV}$,
contains linear terms of $A$ to which small coefficients are attached.
Though divergent at the $A\to\infty$ limit,
these terms should correspond to expansion with respect to $A$.
We plot $U_0^\mathrm{emp}$ and $U_1^\mathrm{emp}$ at $A=208$ and $400$
to view the values at large $A$.

Although the empirical potentials do not match one another precisely,
suggesting ambiguity in the extrapolation,
the qualitative trend is similar.
We point out that $U_0$ at the saturation energy
$\epsilon_N\to \epsilon_0\approx -16\,\mathrm{MeV}$
is constrained by the condition
$\epsilon_0 = k_\mathrm{F}^2/(2M) + U_0(\epsilon_0;\rho_0)$.
It is also noted that the slope of $U_0$ at $\epsilon_0$
corresponds to the effective mass (to be precise, the $k$-mass),
which is constrained by the nuclear structure.
Nevertheless, Fig.~\ref{fig:U_NM} clarifies
that $U_0$ significantly depends on the effective interactions
as $\epsilon_N$ becomes several tens MeV.
In particular, the Gogny-D1S interaction provides $\epsilon_N$-dependence
different from the empirical potentials.
In contrast, $U_0$ with M3Y-P6 resembles
$U_0^\mathrm{emp}$ of the KD potential for $A=400$.
$U_0$ and $U_1$ with the Skyrme interaction are linear functions
of $\epsilon_N$,
and the $k$-mass determines the slope of $U_0$.
$U_0$ from the Skyrme interaction does not severely deviate
from the empirical potential
as long as the $k$-mass has been adjusted as in SLy4,
though it cannot describe a slight bend of $U_0$.

It is hard to constrain $U_1$ from the nuclear structure.
The $\epsilon_N$-dependence of $U_1$
is distinctive among the effective interactions.
M3Y-P6 provides $U_1$ almost consistent with the empirical potentials
and with a microscopic result reported in Ref.~\cite{ref:JLM77},
decreasing almost linearly for growing $\epsilon_N$,
in contrast to SLy4 and D1S.
These properties of M3Y-P6 could originate
from its links to the bare nucleonic interaction.
While the applications of the Skyrme and the Gogny interactions
have been limited to $\epsilon_N\lesssim 30\,\mathrm{MeV}$,
it will deserve testing M3Y-P6 for $N$-$A$ elastic scattering
even at higher energies.

\section{$N$-$A$ scattering cross-sections\label{sec:DCS}}

We now turn to finite nuclei.
As mentioned above,
$U\,(=\mathcal{V}^\mathrm{SFP})$ of Eq.~\eqref{eq:sp-pot_N}
provides a non-local potential, in general.
Because the non-local SFP needs the s.p. wave-functions
beyond the local density $\rho(\mathbf{r})$
and somewhat complicated computation,
the local approximation has customarily been applied~\cite{ref:BR77}.
However, for consistency with the nuclear structure calculations,
we apply the non-local SFP
up to the non-central and Coulomb channels.
Formulae deriving the non-local SFP from the effective interaction
and the MF wave-functions
are given in Appendix~\ref{app:SFP}.
We emphasize that the present non-local SFP
is independent of energy ($\epsilon_N$).
The $\epsilon_N$-dependence of $U_0$ and $U_1$ in Fig.~\ref{fig:U_NM}
results merely from converting the non-locality to the momentum dependence.

In this work, we investigate the $N$-$A$ elastic scattering
at the incident energies ranging
from $\epsilon_N\approx 10\,\mathrm{MeV}$ to $80\,\mathrm{MeV}$.
For the target nuclei,
we pick up $^{16}$O, $^{40}$Ca, $^{90}$Zr and $^{208}$Pb,
which have $J^\pi=0^+$ and whose wave-function can reasonably be obtained
by the spherical Hartree-Fock (HF) calculation.
On top of the self-consistent HF wave-function of the target nucleus
with M3Y-P6,
the wave-function of the scattered nucleon is calculated
under the optical potential,
whose real part is taken from Eq.~\eqref{eq:sp-pot_N}
with the same M3Y-P6 interaction.
For the imaginary part,
we employ the empirical potential $\mathcal{W}^\mathrm{emp}$
of Ref.~\cite{ref:KD03},
which is local and $\epsilon_N$-dependent.
Thus, the optical potential is
$\mathcal{U}=\mathcal{V}^\mathrm{SFP}+i\,\mathcal{W}^\mathrm{emp}$.
We then compute physical quantities with the SIDES code~\cite{ref:SIDES}.
Because the imaginary potential is connected
to the inelastic scattering and the particle emission,
its microscopic description should be consistent with these processes,
and is left for future works.
Whereas we have also tried other empirical imaginary
potentials~\cite{ref:CH89,ref:Sch82,ref:Wat69},
the results are similar to those with the potential of Ref.~\cite{ref:KD03}.
Influence of the center-of-mass (c.m.) Hamiltonian on the $N$-$A$ scattering
is discussed in Appendix.~\ref{app:cm}.
The $[-\mathbf{P}_A^2/(2AM)]$ term in Eq.~\eqref{eq:cmcorr}
has been included in the HF calculations~\cite{ref:Nak20}.
The c.m. correction of the first term on the rhs in Eq.~\eqref{eq:cmcorr}
is handled within the SIDES code~\cite{ref:SIDES}.

\begin{figure}
  \hspace*{-4.5cm}\includegraphics[scale=0.5]{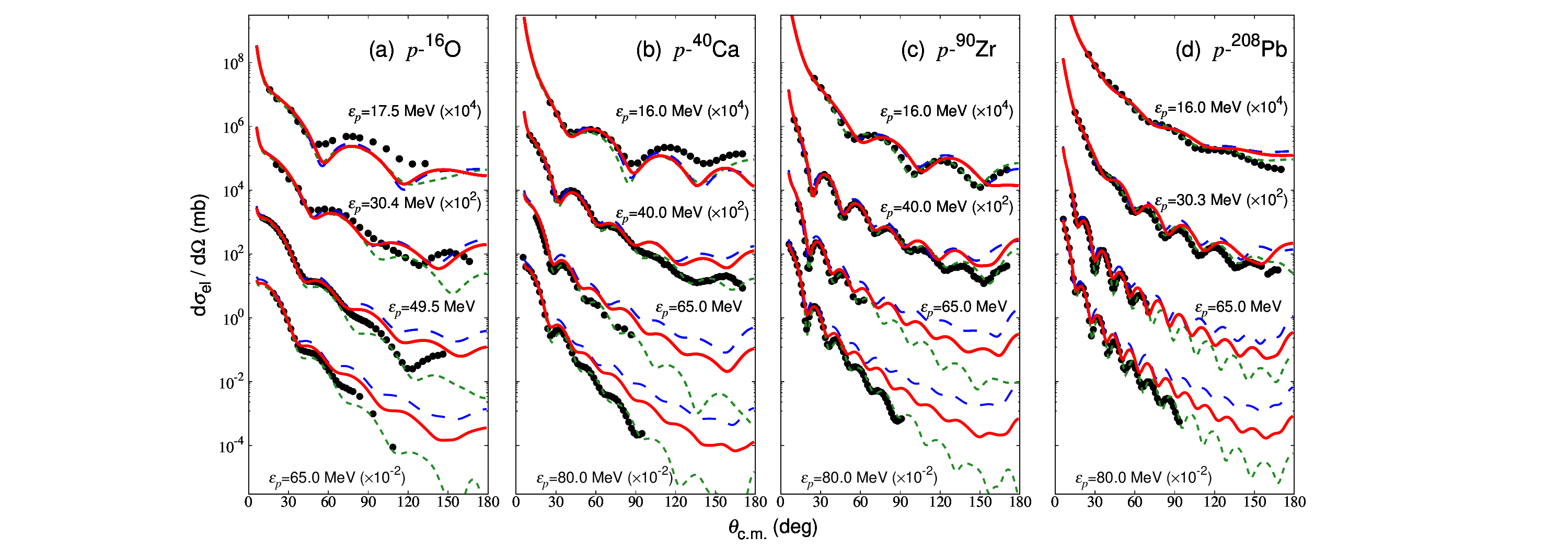}\vspace*{-8mm}
\caption{$d\sigma_\mathrm{el}/d\Omega$ of $p$-$A$ scattering:
  (a) $p$-$^{16}$O, (b) $p$-$^{40}$Ca, (c) $p$-$^{90}$Zr and (d) $p$-$^{208}$Pb.
  Results of $\mathcal{V}^\mathrm{SFP}$ with M3Y-P6
  plus $\mathcal{W}^\mathrm{emp}$ are depicted by red solid lines
  and compared with experimental data (black circles)
  taken from the database~\cite{ref:EXFOR},
  originally reported in Refs.~\protect\cite{ref:CH89,ref:exp_DCS_RCNP,
        ref:exp_DCS_p-O16_e17,ref:exp_DCS_p-O16_e30,ref:exp_DCS_p-O16_e50,
        ref:exp_DCS_p-A_e40,ref:exp_DCS_p-A_e80,ref:exp_DCS_p-Pb208}.
  For comparison,
  results of $\mathcal{V}^\mathrm{SFP}$ with Gogny-D1S (blue long-dashed line),
  and those of $\mathcal{V}^\mathrm{emp}$ 
    of Ref.~\protect\cite{ref:KD03}
  (green short-dashed lines)
  are also displayed.
  Depending on $\epsilon_p$,
  the cross sections are scaled by the coefficients in the parentheses.
\label{fig:p-A_dsig}}
\end{figure}

In Fig.~\ref{fig:p-A_dsig},
the calculated differential cross-sections $d\sigma_\mathrm{el}/d\Omega$
of the proton-nucleus ($p$-$A$) scatterings
are compared with the experimental data~\cite{ref:EXFOR}.
As well as the results of $\mathcal{V}^\mathrm{SFP}$ with M3Y-P6,
we display the results with the Gogny-D1S interaction~\cite{ref:D1S},
and those applying the phenomenological potential of Ref.~~\cite{ref:KD03}
also to the real part, $\mathcal{V}^\mathrm{emp}$.
Covering light to heavy nuclei
ranging from $\epsilon_p\approx 15\,\mathrm{MeV}$ to $80\,\mathrm{MeV}$,
the SFPs with M3Y-P6 reproduce $d\sigma_\mathrm{el}/d\Omega$ well,
almost comparably to the empirical potential
but without adjusting $\mathcal{V}$ to the scattering data.
In particular, notable agreement with the data
is found at $\epsilon_p=65\,\mathrm{MeV}$.
At $\epsilon_p=80\,\mathrm{MeV}$,
the calculated $d\sigma_\mathrm{el}/d\Omega$ is larger than the data
at $\theta_\mathrm{c.m.}\gtrsim 60^\circ$.
Still, the positions of the peaks and dips are well reproduced.
In contrast, the D1S interaction gives $d\sigma_\mathrm{el}/d\Omega$
seriously deviating from the data in $\epsilon_p\gtrsim 65\mathrm{MeV}$,
while it reproduces the cross sections
at $\epsilon_p\lesssim 30\,\mathrm{MeV}$.
This seems connected with the $\epsilon_N$-dependence of $U_0$
in Fig.~\ref{fig:U_NM}.

\begin{figure}
  \includegraphics[scale=0.6]{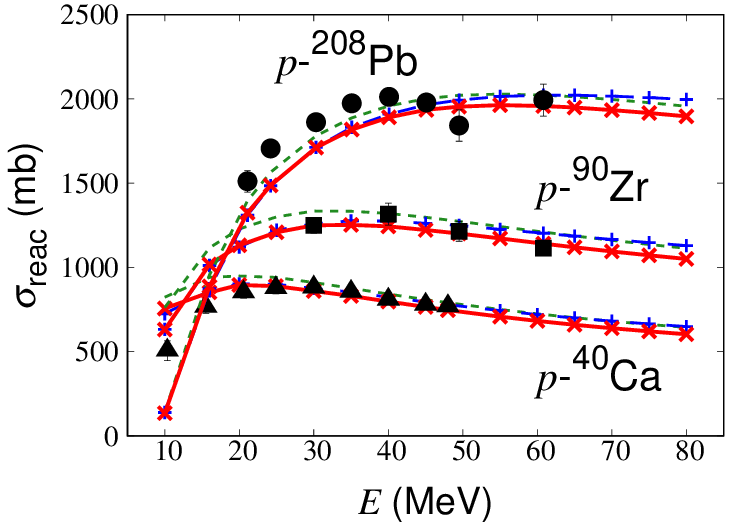}\vspace*{-8mm}
  \caption{$\sigma_\mathrm{reac}$ of $p$-$^{40}$Ca, $p$-$^{90}$Zr
    and $p$-$^{208}$Pb scattering.
    Results of $\mathcal{V}^\mathrm{SFP}$ with M3Y-P6 (D1S)
    plus $\mathcal{W}^\mathrm{emp}$ are plotted
    by crosses connected with red solid lines
    (pluses with blue long-dashed lines),
    and those of $\mathcal{V}^\mathrm{emp}$ are displayed
    by green short-dashed lines.
    Black circles, squares and triangles
    are experimental data~\protect\cite{ref:EXFOR,ref:exp_p-RCS1,
    ref:exp_p-RCS2,ref:exp_p-RCS3}.
    \label{fig:p-A_sig_R}}
\end{figure}

The optical theorem gives
the total cross-section $\sigma_\mathrm{tot}$~\cite{ref:Sat83},
from which the reaction cross-section $\sigma_\mathrm{reac}$ is obtained as
\begin{equation}
  \sigma_\mathrm{reac} = \sigma_\mathrm{tot}
  - \int d\Omega\,\frac{d\sigma_\mathrm{el}}{d\Omega}\,.
\label{eq:sig_R}\end{equation}
While both terms on the rhs are divergent in the $p$-$A$ scatterings,
$\sigma_\mathrm{reac}$ is calculated in the SIDES code
by properly treating the Coulombic contribution
as discussed in Ref.~\cite{ref:Sat83}.
Although $\sigma_\mathrm{reac}$ is primarily subject to the imaginary potential,
the real potential indirectly influences $\sigma_\mathrm{reac}$.
We show $\sigma_\mathrm{reac}$'s in the $p$-$A$ scatterings
in Fig.~\ref{fig:p-A_sig_R}
to examine the consistency of $\mathcal{W}^\mathrm{emp}$
combined with $\mathcal{V}^\mathrm{SFP}$.
As expected, $\sigma_\mathrm{reac}$'s are insensitive to $\mathcal{V}$,
and the application of $\mathcal{W}^\mathrm{emp}$
in combination with $\mathcal{V}^\mathrm{SFP}$ is justified
in $10\lesssim \epsilon_p\lesssim 65\,\mathrm{MeV}$.
Without available data,
there remains room to improve $d\sigma_\mathrm{el}/d\Omega$
by readjusting $\mathcal{W}$ at $\epsilon_p\approx 80\,\mathrm{MeV}$,
although an upper limit in $\epsilon_N$ is anticipated
for the applicability of $\mathcal{V}^\mathrm{SFP}$,
as argued below.

\begin{figure}
  \includegraphics[scale=0.5]{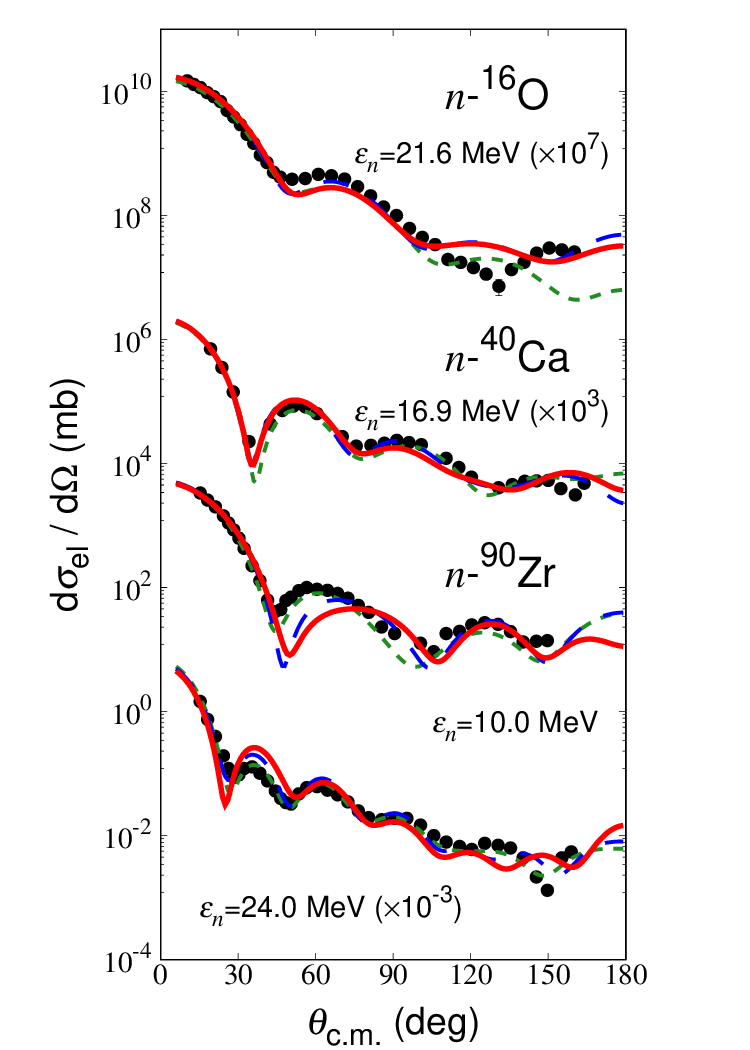}\vspace*{-8mm}
  \caption{$d\sigma_\mathrm{el}/d\Omega$ of $n$-$A$ scattering:
    $n$-$^{16}$O, $n$-$^{40}$Ca and $n$-$^{90}$Zr.
    Results of $\mathcal{V}^\mathrm{SFP}$ with M3Y-P6 and D1S,
    and those of $\mathcal{V}^\mathrm{emp}$ are presented,
    in comparison with experimental data~\protect~\cite{ref:EXFOR,
      ref:exp_DCS_n-O16,
        ref:exp_DCS_n-Ca40,ref:exp_DCS_n-Zr90}.
    See Fig.~\protect\ref{fig:p-A_dsig} for conventions.
\label{fig:n-A_dsig}}
\end{figure}

The calculated $d\sigma_\mathrm{el}/d\Omega$
of the neutron-nucleus ($n$-$A$) scatterings
at $\epsilon_n\lesssim 30\,\mathrm{MeV}$
are compared with the experimental data~\cite{ref:EXFOR}
in Fig.~\ref{fig:n-A_dsig}.
Data at higher energies are limited to small angles.
It is confirmed that the calculated $\sigma_\mathrm{tot}$'s are compatible
with the data in this energy range.

Many effective interactions developed for scattering
have explicit energy ($\epsilon$) dependence in their parameters,
connected to the $\epsilon$-dependence of the $G$-matrix.
However, the $\epsilon$-dependent interaction complicates
treating the nuclear structure and finite-$T$ problems,
in which a single system includes s.p. states with various energies.
Therefore, $\epsilon$-independent effective interactions
appropriately containing correlation effects are valuable.
Still, it would be too optimistic to believe
that we can remove $\epsilon$ dependence everywhere.
There will be an upper limit of $\epsilon$
where the $\epsilon$-independent interaction works.
It seems reasonable to consider that, for the present M3Y-P6 interaction,
the upper limit lies around $\epsilon=80\,\mathrm{MeV}$.

In this paper, we have yet to discuss the analyzing power,
on which some experimental data are available.
The analyzing power is primarily relevant to the non-central channels,
which do not contribute to the energy in homogeneous matter.
We have confirmed that influence of the non-central channels,
which are included except in the calculations for Fig.~\ref{fig:U_NM},
is insignificant for $d\sigma_\mathrm{el}/d\Omega$.
The analyzing power will be argued in a forthcoming paper.

\section{Summary and outlook\label{sec:summary}}

Based on the variational principle,
we have discussed the self-consistent s.p. potential
at positive energy.
As it corresponds to the SFP
and thereby to the real part of the optical potential,
the property and validity of the potential
can be examined via the $N$-$A$ elastic scattering.

For homogeneous nuclear matter,
the non-locality of the potential is convertible to the $\epsilon$ dependence,
where $\epsilon$ corresponds to the energy of the incident nucleon
in the $N$-$A$ elastic scattering.
We calculate the nuclear-matter s.p. potential
with several effective interactions that work well for nuclear structure,
and compare them to the empirical optical potentials at large $A$.
It is found that some of them do not reproduce
the $\epsilon$ dependence of the empirical potential.
We have shown that the semi-realistic nucleonic interaction M3Y-P6,
which has links to the bare nucleonic interaction,
provides s.p. potentials similar to the empirical ones.

We have calculated the real part of the SFP
fully consistent with the nuclear structure calculations.
The differential cross-sections of the $N$-$A$ elastic scatterings
have been computed by applying M3Y-P6
both to the SCMF calculations for the target nucleus
and to the scattered nucleon's SFP.
The imaginary part of the optical potential,
which carries the absorption effects,
is complemented by the empirical one.
It is confirmed that the SFP with M3Y-P6 describes
the cross sections well,
almost comparable to the empirical potentials,
in as wide an energy range as $\epsilon\lesssim 80\,\mathrm{MeV}$.
The reaction cross-sections of the proton scatterings
and the total cross-sections of the neutron scatterings
do not contradict the elastic scattering results.

United with the already established nuclear structure results,
the present results demonstrate
that s.p. potentials compatible with experimental data
can be derived from a single $\epsilon$-independent effective interaction
in a significantly extended energy range,
indicating that the effective interaction
properly takes account of many-body correlations.
As the effective interaction is the only input in the SCMF (or KS) approach,
these results may be a yardstick
for extending the SCMF approach in nuclei to finite $T$
without artificial discontinuity with respect to energy.
\appendix

\section{Effective Hamiltonian\label{app:hamil}}

In the subsequent Appendices,
we consider the following Hamiltonian
for the system comprised of $A'$ nucleons,
\begin{equation}\begin{split} H =& K + V_N + V_C - H_\mathrm{c.m.}\,;\\
& K = \sum_\alpha \frac{\mathbf{p}_\alpha^2}{2M}\,,\quad
V_N = \sum_{\alpha<\beta} v_{\alpha\beta}\,,\quad
V_C = \alpha_\mathrm{em} \sum_{\alpha<\beta(\in p)} \frac{1}{r_{\alpha\beta}}\,,\\
& H_\mathrm{c.m.} = \frac{\mathbf{P}^2}{2A'M}
= \frac{1}{A'}\bigg[\sum_\alpha \frac{\mathbf{p}_\alpha^2}{2M}
  + \sum_{\alpha<\beta} \frac{\mathbf{p}_\alpha\cdot\mathbf{p}_\beta}{M}\bigg]\quad
\Big(\mathbf{P}=\sum_\alpha \mathbf{p}_\alpha\Big)\,,
\end{split}\label{eq:Hamil}\end{equation}
where $\mathbf{r}_{\alpha\beta}= \mathbf{r}_\alpha - \mathbf{r}_\beta$
with $r=|\mathbf{r}|$,
and $\alpha_\mathrm{em}$ (in $V_C$) is the fine structure constant.
The effective nucleonic interaction $v_{\alpha\beta}$
is comprised of the following terms,
\begin{equation}\begin{split} v_{\alpha\beta} =& v_{\alpha\beta}^{(\mathrm{C})}
 + v_{\alpha\beta}^{(\mathrm{LS})} + v_{\alpha\beta}^{(\mathrm{TN})}
 + v_{\alpha\beta}^{(\mathrm{C}\rho)} + v_{\alpha\beta}^{(\mathrm{LS}\rho)}\,;\\
& v_{\alpha\beta}^{(\mathrm{C})} = \sum_k \big\{t_k^{(\mathrm{SE})} P_\mathrm{SE}
+ t_k^{(\mathrm{TE})} P_\mathrm{TE} + t_k^{(\mathrm{SO})} P_\mathrm{SO}
+ t_k^{(\mathrm{TO})} P_\mathrm{TO}\big\}\,
 f_k^{(\mathrm{C})} (r_{\alpha\beta})\,,\\
& v_{\alpha\beta}^{(\mathrm{LS})} = \sum_k \big\{t_k^{(\mathrm{LSE})} P_\mathrm{TE}
 + t_k^{(\mathrm{LSO})} P_\mathrm{TO}\big\}\,
 f_k^{(\mathrm{LS})} (r_{\alpha\beta})\,\mathbf{L}_{\alpha\beta}\cdot
(\mathbf{s}_\alpha+\mathbf{s}_\beta)\,,\\
& v_{\alpha\beta}^{(\mathrm{TN})} = \sum_k \big\{t_k^{(\mathrm{TNE})} P_\mathrm{TE}
 + t_k^{(\mathrm{TNO})} P_\mathrm{TO}\big\}\,
 f_k^{(\mathrm{TN})} (r_{\alpha\beta})\, r_{\alpha\beta}^2 S_{\alpha\beta}\,,\\
 & v_{\alpha\beta}^{(\mathrm{C}\rho)}
 = \Big\{C_\mathrm{SE}[\rho(\mathbf{R}_{\alpha\beta})]\,P_\mathrm{SE}
 + C_\mathrm{TE}[\rho(\mathbf{R}_{\alpha\beta})]\,P_\mathrm{TE}\Big\}\,
 \delta(\mathbf{r}_{\alpha\beta})\,,
 \\
& v_{\alpha\beta}^{(\mathrm{LS}\rho)} = 2i\,D[\rho(\mathbf{R}_{\alpha\beta})]\,
 \mathbf{p}_{\alpha\beta}\times\delta(\mathbf{r}_{\alpha\beta})\,
 \mathbf{p}_{\alpha\beta}\cdot(\mathbf{s}_\alpha+\mathbf{s}_\beta) \\
 &\qquad~ = D[\rho(\mathbf{R}_{\alpha\beta})]\,\{-\nabla_{\alpha\beta}^2\,
 \delta(\mathbf{r}_{\alpha\beta})\}\,
 \mathbf{L}_{\alpha\beta}\cdot(\mathbf{s}_\alpha+\mathbf{s}_\beta)\,,
\end{split}\label{eq:effint}\end{equation}
where $\mathbf{s}_\alpha$ is the spin operator,
$\mathbf{R}_{\alpha\beta}=(\mathbf{r}_\alpha+\mathbf{r}_\beta)/2$,
$\mathbf{p}_{\alpha\beta}= (\mathbf{p}_\alpha - \mathbf{p}_\beta)/2$,
$\mathbf{L}_{\alpha\beta}= \mathbf{r}_{\alpha\beta}\times \mathbf{p}_{\alpha\beta}$,
$S_{\alpha\beta}= 4\,[3(\mathbf{s}_\alpha\cdot\hat{\mathbf{r}}_{\alpha\beta})
(\mathbf{s}_\beta\cdot\hat{\mathbf{r}}_{\alpha\beta})
- \mathbf{s}_\alpha\cdot\mathbf{s}_\beta ]$,
and $\rho(\mathbf{r})$ is the isoscalar nucleon density.
$P_\mathrm{Y}$ ($\mathrm{Y}=\mathrm{SE},\mathrm{TE},\mathrm{SO},\mathrm{TO}$)
stands for the projection operators on the singlet-even (SE), triplet-even (TE),
singlet-odd (SO) and triplet-odd (TO) two-nucleon states.
They are related to the spin- and isospin-exchange operators
$P_\sigma\,[=(1+4\mathbf{s}_\alpha\cdot\mathbf{s}_\beta)/2]$ and $P_\tau$ as
\begin{equation}\begin{split}
P_\mathrm{SE} = \frac{1-P_\sigma}{2}\,\frac{1+P_\tau}{2}\,,
\quad& P_\mathrm{TE} = \frac{1+P_\sigma}{2}\,\frac{1-P_\tau}{2}\,,\\
P_\mathrm{SO} = \frac{1-P_\sigma}{2}\,\frac{1-P_\tau}{2}\,.
\quad& P_\mathrm{TO} = \frac{1+P_\sigma}{2}\,\frac{1+P_\tau}{2}\,.
\end{split}\label{eq:proj_T}\end{equation}
Each channel $\mathrm{X}\,(=\mathrm{C},\mathrm{LS},\mathrm{TN}$)
is composed of several terms distinguished by $k$,
which corresponds to the function $f_k^{(\mathrm{X})}(r)$
and contains coupling constants $t_k^{(\mathrm{Y})}$.
In the M3Y-type interaction,
the Yukawa function $f_k^{(\mathrm{X})}(r)=e^{-x_k}/x_k$
with $x_k=\mu_k^{(\mathrm{X})} r$
is used for all of $\mathrm{X}=\mathrm{C},\mathrm{LS},\mathrm{TN}$,
where $\mu_k^{-1}$ is the interaction range.
In the conventional Gogny interaction,
$f_k^{(\mathrm{C})}(r)=e^{-(\mu_k^{(\mathrm{C})} r)^2}$ and
$f^{(\mathrm{LS})}(r)=\nabla^2\delta(\mathbf{r})$.
The expression \eqref{eq:effint} also covers the Skyrme interaction
by setting $f_1^{(\mathrm{C})}(r)=\delta(\mathbf{r})$,
$f_2^{(\mathrm{C})}(r)=f^{(\mathrm{LS})}(r)=f^{(\mathrm{TN})}(r)
=\nabla^2\delta(\mathbf{r})$~\cite{ref:NS02}.
The $v^{(\mathrm{C}\rho)}$ and $v^{(\mathrm{LS}\rho)}$ terms
have coupling constants $C_\mathrm{Y}[\rho]$ and $D[\rho]$
that depend on $\rho$,
whose functional forms need not be specified here.

For later convenience,
we rewrite $v_{\alpha\beta}^{(\mathrm{X})}$
($\mathrm{X}=\mathrm{C},\mathrm{LS},\mathrm{TN}$) as
\begin{equation}\begin{split}
  v_{\alpha\beta}^{(\mathrm{C})} =& \sum_k \Big[\big\{\bar{t}_k^{(0\mathrm{i})}
  + \bar{t}_k^{(0\mathrm{x})} P_\tau\big\}
  + (4\mathbf{s}_\alpha\cdot\mathbf{s}_\beta)\,\big\{\bar{t}_k^{(1\mathrm{i})}\,
  + \bar{t}_k^{(1\mathrm{x})}\,P_\tau\big\}\Big]\,
  f_k^{(\mathrm{C})} (r_{\alpha\beta})\,;\\
  &\qquad \bar{t}_k^{(0\mathrm{i})}
  = \frac{1}{8}\big(t_k^{(\mathrm{SE})}+3t_k^{(\mathrm{TE})}
  +t_k^{(\mathrm{SO})}+3t_k^{(\mathrm{TO})}\big)\,,\\
  &\qquad \bar{t}_k^{(0\mathrm{x})}
  = \frac{1}{8}\big(t_k^{(\mathrm{SE})}-3t_k^{(\mathrm{TE})}
  -t_k^{(\mathrm{SO})}+3t_k^{(\mathrm{TO})}\big)\,,\\
  &\qquad \bar{t}_k^{(1\mathrm{i})}
  = \frac{1}{8}\big(-t_k^{(\mathrm{SE})}+t_k^{(\mathrm{TE})}
  -t_k^{(\mathrm{SO})}+t_k^{(\mathrm{TO})}\big)\,,\\
  &\qquad \bar{t}_k^{(1\mathrm{x})}
  = \frac{1}{8}\big(-t_k^{(\mathrm{SE})}-t_k^{(\mathrm{TE})}
  +t_k^{(\mathrm{SO})}+t_k^{(\mathrm{TO})}\big)\,,\\
  v_{\alpha\beta}^{(\mathrm{LS})} =& \sum_k \big\{\bar{t}_k^{(\mathrm{LSi})}
  + \bar{t}_k^{(\mathrm{LSx})} P_\tau\big\}\,f_k^{(\mathrm{LS})} (r_{\alpha\beta})\,
  \mathbf{L}_{\alpha\beta}\cdot(\mathbf{s}_\alpha+\mathbf{s}_\beta)\,;\\
  &\qquad \bar{t}_k^{(\mathrm{LSi})} = \frac{1}{2}
  \big(t_k^{(\mathrm{LSE})}+t_k^{(\mathrm{LSO})}\big)\,,\quad
  \bar{t}_k^{(\mathrm{LSx})} = \frac{1}{2}
  \big(-t_k^{(\mathrm{LSE})}+t_k^{(\mathrm{LSO})}\big)\,,\\
  v_{\alpha\beta}^{(\mathrm{TN})} =& \sum_k \big\{\bar{t}_k^{(\mathrm{TNi})}
  + \bar{t}_k^{(\mathrm{TNx})} P_\tau\big\}\,f_k^{(\mathrm{TN})} (r_{\alpha\beta})\,
  r_{\alpha\beta}^2 S_{\alpha\beta}\,;\\
  &\qquad \bar{t}_k^{(\mathrm{TNi})} = \frac{1}{2}
  \big(t_k^{(\mathrm{TNE})}+t_k^{(\mathrm{TNO})}\big)\,,\quad
  \bar{t}_k^{(\mathrm{TNx})} = \frac{1}{2}
  \big(-t_k^{(\mathrm{TNE})}+t_k^{(\mathrm{TNO})}\big)\,.
\end{split}\label{eq:eff_rewrite}\end{equation}
The LS and tensor operators are expanded as,
using $\boldsymbol{\ell}=\mathbf{r}\times\mathbf{p}$,
\begin{equation}\begin{split}
    \mathbf{L}_{\alpha\beta}\cdot(\mathbf{s}_\alpha+\mathbf{s}_\beta)
    &= \frac{1}{2}\big\{\boldsymbol{\ell}_\alpha\cdot\mathbf{s}_\alpha
    + \boldsymbol{\ell}_\alpha\cdot\mathbf{s}_\beta
    + \boldsymbol{\ell}_\beta\cdot\mathbf{s}_\alpha
    + \boldsymbol{\ell}_\beta\cdot\mathbf{s}_\beta
    + (\mathbf{r}_\alpha\times\mathbf{s}_\alpha)\cdot\mathbf{p}_\beta
    - (\mathbf{p}_\alpha\times\mathbf{s}_\alpha)\cdot\mathbf{r}_\beta \\
    &\qquad\qquad - \mathbf{r}_\alpha\cdot(\mathbf{p}_\beta\times\mathbf{s}_\beta)
    + \mathbf{p}_\alpha\cdot(\mathbf{r}_\beta\times\mathbf{s}_\beta)\big\}\,,\\
    r_{\alpha\beta}^2 S_{\alpha\beta}
    =& 16\pi \sum_{\lambda_\alpha \lambda_\beta}
    \delta_{\lambda_\alpha+\lambda_\beta,2}\,\zeta_{\lambda_\alpha}\,
    r_\alpha^{\lambda_\alpha} r_\beta^{\lambda_\beta}\,
    \big[Y^{(\lambda_\alpha)}(\hat{\mathbf{r}}_\alpha)\,
      Y^{(\lambda_\beta)}(\hat{\mathbf{r}}_\beta)\big]^{(2)}\cdot
    [s_\alpha^{(1)} s_\beta^{(1)}]^{(2)}\,;\\
    &\qquad\qquad \zeta_0=\zeta_2=\sqrt{\frac{6}{5}}\,,~\zeta_1=-2\,.
\end{split}\end{equation}
  
\section{Formulae for single folding potential\label{app:SFP}}

This Appendix provides formulae to obtain the single folding potential (SFP)
from the effective interaction in Appendix~\ref{app:hamil}.
The wave function (w.f.) of the target nucleus is assumed to be obtained
by a Hartree-Fock (HF) calculation.

In deriving the formulae of the SFP,
we express the w.f. of the target nucleus
via a set of the s.p. basis functions,
which are denoted by $\varphi_\alpha(\mathbf{r}\sigma\tau)$,
with the spin and isospin coordinates $\sigma$ and $\tau\,(=p,n)$.
Instead of the occupation probability $n^{(0)}_\alpha$ in the text,
we employ the one-body density matrix $\varrho_{\alpha\alpha'}
(:=\langle\Phi|a^\dagger_{\alpha'} a_\alpha|\Phi\rangle)$.
The w.f. of the projectile nucleon is represented
by $\psi(\mathbf{r}\sigma\tau)$.
The energy of the whole $N$-$A$ system is composed
of the individual terms of the effective Hamiltonian.
The SFP is expressed by a sum of the terms
corresponding to those in Eq.~\eqref{eq:effint},
\begin{equation}\begin{split}
  U|N\rangle &= \sum_\mathrm{X} U^{(\mathrm{X})}|N\rangle \\
  &= \sum_\mathrm{X} \bigg\{
  \sum_{\alpha \alpha'\in A} \langle\ast\alpha'|v^\mathrm{(X)}
  |N\alpha\rangle\,\varrho_{\alpha\alpha'}
  + \frac{1}{2}\,|N\rangle\,\sum_{\alpha \alpha' \beta \beta'\in A}
  \big\langle \alpha'\beta'\big|
  \frac{\delta v^\mathrm{(X)}}{\delta\langle N|}
  \big|\alpha\beta\big\rangle\,\varrho_{\alpha\alpha'}\,\varrho_{\beta\beta'}
  \bigg\}\,.
\end{split}\label{eq:U_tot}\end{equation}
See the text for the notation.
For density-independent channels
($\mathrm{X}=\mathrm{C},\mathrm{LS},\mathrm{TN}$),
we have no rearrangement term,
and Eq.~\eqref{eq:U_tot} yields
\begin{equation}\begin{split}
  U^{(\mathrm{X})} =& U^{(\mathrm{X,dir})} + U^{(\mathrm{X,exc})}\,;\\
  &\quad \big[U^{(\mathrm{X,dir})} \psi\big](\mathbf{r}\sigma\tau)
  = \sum_{\alpha \alpha'\in A} \varrho_{\alpha\alpha'}
  \sum_{\sigma'\tau'} \int d^3r'\,\varphi_{\alpha'}^\ast(\mathbf{r}'\sigma'\tau')\,
  v^{(\mathrm{X})}\,
  \varphi_\alpha(\mathbf{r}'\sigma'\tau')\,\psi(\mathbf{r}\sigma\tau)\,,\\
  &\quad \big[U^{(\mathrm{X,exc})} \psi\big](\mathbf{r}\sigma\tau)
  = - \sum_{\alpha \alpha'\in A} \varrho_{\alpha\alpha'}
  \sum_{\sigma'\tau'} \int d^3r'\,\varphi_{\alpha'}^\ast(\mathbf{r}'\sigma'\tau')\,
  v^{(\mathrm{X})}\,
    \varphi_\alpha(\mathbf{r}\sigma\tau)\,\psi(\mathbf{r}'\sigma'\tau')\,.
\end{split}\label{eq:U_X}\end{equation}
The Coulomb interaction can be handled analogously.
The density-dependent channels of Eq.~\eqref{eq:U_tot}
($\mathrm{X}=\mathrm{C}\rho,\mathrm{LS}\rho$) will be discussed
without separating the direct and exchange terms
since they are assumed to be zero range,
though the density-dependence leads to the rearrangement term
(the second term on the rhs).

In the following,
we omit the subscript $N$ for the projectile unless it leads to confusion.
Furthermore, we drop the subscript $k$
that distinguishes the range parameters in Eq.~\eqref{eq:effint},
and the summation over $k$,
though each channel may include plural terms having different ranges.
Spherical basis functions are adopted for $\varphi_\alpha$
and the partial-wave expansion is applied to $\psi$,
\begin{equation}\begin{split}
    \varphi_\alpha(\mathbf{r}\sigma\tau) &= \delta_{\tau\tau_\alpha}
    R_{\nu_\alpha\ell_\alpha j_\alpha}(r)\,
    [Y^{(\ell_\alpha)}(\hat{\mathbf{r}})\,\chi_\sigma^{(1/2)}]^{(j_\alpha)}_{m_\alpha}\,
    \xi_\tau\,,\\
    \psi(\mathbf{r}\sigma\tau_N) &= \delta_{\tau\tau_N}
    \sum_{\ell j m} c_{\ell j m} \mathcal{R}_{\ell j}(r)\,
        [Y^{(\ell)}(\hat{\mathbf{r}})\,\chi_\sigma^{(1/2)}]^{(j)}_m\,\xi_\tau\,,
\end{split}\label{eq:basis}\end{equation}
where $\chi_\sigma$ ($\xi_\tau$) is the spin (isospin) w.f.
and $c_{\ell j m}$ is an appropriate coefficient.
If the target has $J^\pi=0^+$,
the density matrix has the property $\varrho_{\alpha\alpha'}
= \delta_{\tau_\alpha \tau_{\alpha'}}\delta_{\ell_\alpha \ell_{\alpha'}}
\delta_{j_\alpha j_{\alpha'}}\delta_{m_\alpha m_{\alpha'}}\,
\varrho^{(\tau_\alpha \ell_\alpha j_\alpha)}_{\nu_\alpha \nu_{\alpha'}}$.
The quantum numbers $(\ell j m)$ do not mix in $\psi$,
with $c_{\ell j m}$ fixed by the incident wave.
The SFP is obtained for each $(\ell j)$,
which will be denoted by $U_{\ell j}^{(\mathrm{X})}$.

The function $f(r_{\alpha\beta})$ in Eq.~\eqref{eq:effint} is expanded as
\begin{equation}
  f(r_{\alpha\beta})
  = \sum_\lambda g_\lambda(r_\alpha,r_\beta)\,
  P_\lambda(\hat{\mathbf{r}}_\alpha\cdot\hat{\mathbf{r}}_\beta)
  = \sum_\lambda \frac{4\pi}{2\lambda+1}\,
  g_\lambda(r_\alpha,r_\beta)\,Y^{(\lambda)}(\hat{\mathbf{r}}_\alpha)\cdot
  Y^{(\lambda)}(\hat{\mathbf{r}}_\beta)\,,
\end{equation}
where $P_\lambda$ is the Legendre polynomial and
\begin{equation}
  g_\lambda(r_\alpha,r_\beta)
  = \frac{2\lambda+1}{2}\,\int_{-1}^1 d(\cos\theta)\,
  f\Big(\sqrt{r_\alpha^2+r_\beta^2-2r_\alpha r_\beta\cos\theta}\,\Big)\,
  P_\lambda(\cos\theta)\,.
\label{eq:f_lambda}\end{equation}
The form of $g_\lambda$ for several functions will be given later.

\subsection{Terms from central channels\label{subapp:central}}

The contribution of $v^{(\mathrm{C})}$ to the SFP
for each $(\ell, j)$ partial wave is represented as
\begin{equation}\begin{split}
  U^{(\mathrm{C,dir})}_{\ell j} &= \sum_{\gamma=0,1} \sum_{\tau_\alpha}
  \big\{\bar{t}^{(\gamma\mathrm{i})}
  +\delta_{\tau\tau_\alpha}\bar{t}^{(\gamma\mathrm{x})}\big\}\,
  F^{(\mathrm{dir},\gamma)}_{\ell j,\tau_\alpha}\,;\\
  &\quad F^{(\mathrm{dir},\gamma)}_{\ell j,\tau_\alpha}
  = \sum_{\substack{\ell_\alpha j_\alpha m_\alpha\\ \nu_\alpha \nu_{\alpha'}(\in A)}}
  \varrho^{(\tau_\alpha \ell_\alpha j_\alpha)}_{\nu_\alpha \nu_{\alpha'}}
  \sum_{\sigma'\tau'} \int d^3r'\,\big\{R_{\nu_{\alpha'}\ell_\alpha j_\alpha}(r')\,
  [Y^{(\ell_\alpha)}(\hat{\mathbf{r}}')\,\chi_{\sigma'}^{(1/2)}]^{(j_\alpha)}_{m_\alpha}
   \big\}^\ast\\
   &\hspace*{5cm} \times f^{(\mathrm{C})}\big(|\mathbf{r}-\mathbf{r}'|\big)\,
   \mathcal{O}^{(\mathrm{C},\gamma)}_\sigma\,
   \big\{R_{\nu_\alpha\ell_\alpha j_\alpha}(r')\,
   [Y^{(\ell_\alpha)}(\hat{\mathbf{r}}')\,\chi_{\sigma'}^{(1/2)}]^{(j_\alpha)}_{m_\alpha}
   \big\}\,,\\
  U^{(\mathrm{C,exc})}_{\ell j} &= - \sum_{\gamma=0,1} \sum_{\tau_\alpha}
  \big\{\bar{t}^{(\gamma\mathrm{x})}
  +\delta_{\tau\tau_\alpha}\bar{t}^{(\gamma\mathrm{i})}\big\}\,
  F^{(\mathrm{exc},\gamma)}_{\ell j,\tau_\alpha}\,;\\
  &\quad F^{(\mathrm{exc},\gamma)}_{\ell j,\tau_\alpha}\,
  \big\{\mathcal{R}_{\ell j}(r)\,
        [Y^{(\ell)}(\hat{\mathbf{r}})\,\chi_\sigma^{(1/2)}]^{(j)}_m\big\} \\
  &\qquad = \sum_{\substack{\ell_\alpha j_\alpha m_\alpha\\
        \nu_\alpha \nu_{\alpha'}(\in A)}}
  \varrho^{(\tau_\alpha \ell_\alpha j_\alpha)}_{\nu_\alpha \nu_{\alpha'}}
  \sum_{\sigma'\tau'} \int d^3r'\,\big\{R_{\nu_{\alpha'}\ell_\alpha j_\alpha}(r')\,
  [Y^{(\ell_\alpha)}(\hat{\mathbf{r}}')\,\chi_{\sigma'}^{(1/2)}]^{(j_\alpha)}_{m_\alpha}
  \big\}^\ast\,f^{(\mathrm{C})}\big(|\mathbf{r}-\mathbf{r}'|\big)\,
   \mathcal{O}^{(\mathrm{C},\gamma)}_\sigma\\  
   &\hspace*{5cm} \times
   \big\{R_{\nu_\alpha\ell_\alpha j_\alpha}(r)\,
   [Y^{(\ell_\alpha)}(\hat{\mathbf{r}})\,\chi_{\sigma}^{(1/2)}]^{(j_\alpha)}_{m_\alpha}
   \big\}\,\big\{\mathcal{R}_{\ell j}(r')\,
        [Y^{(\ell)}(\hat{\mathbf{r}}')\,\chi_{\sigma'}^{(1/2)}]^{(j)}_m\big\}\,,\\
   &\hspace*{8cm}(\mathcal{O}^{(\mathrm{C},0)}_\sigma=1\,,~
   \mathcal{O}^{(\mathrm{C},1)}_\sigma=4\mathbf{s}\cdot\mathbf{s}')\,.
\end{split}\label{eq:U_C}\end{equation}
While $F^{(\mathrm{dir},\gamma)}_{\ell j,\tau_\alpha}$ provides a local potential,
$F^{(\mathrm{exc},\gamma)}_{\ell j,\tau_\alpha}$ is an integral operator
whose kernel corresponds to a non-local SFP.
It acts on $\mathcal{R}_{\ell j}$
without influencing the angular-spin function.
The effect of $F^{(\mathrm{exc},\gamma)}_{\ell j,\tau_\alpha}$ becomes transparent 
by integrating out the angular-spin part,
\begin{equation}
    F^{(\mathrm{exc},\gamma)}_{\ell j,\tau_\alpha}\,\mathcal{R}_{\ell j}(r)
    = \sum_\sigma \int d\Omega\,
    \big\{[Y^{(\ell)}(\hat{\mathbf{r}})\,\chi_\sigma^{(1/2)}]^{(j)}_m\big\}^\ast\,
  F^{(\mathrm{exc},\gamma)}_{\ell j,\tau_\alpha}\,\big\{\mathcal{R}_{\ell j}(r)\,
        [Y^{(\ell)}(\hat{\mathbf{r}})\,\chi_\sigma^{(1/2)}]^{(j)}_m\big\}\,,
\label{eq:F_exc}\end{equation}
where $\int d\Omega$ is the integration over the solid angle.
The Racah algebra to the angular-spin part yields
\begin{equation}\begin{split}
    F^{(\mathrm{dir},\gamma)}_{\ell j,\tau_\alpha}
    =&\,\delta_{\gamma 0} \sum_{\substack{\ell_\alpha j_\alpha\\
    \nu_\alpha \nu_{\alpha'}(\in A)}}
    (2j_\alpha+1)\,\varrho^{(\tau_\alpha \ell_\alpha j_\alpha)}_{\nu_\alpha \nu_{\alpha'}}
    \int r^{\prime\,2}dr'\,g_0(r,r')\,R_{\nu_{\alpha'}\ell_\alpha j_\alpha}^\ast(r')\,
    R_{\nu_\alpha \ell_\alpha j_\alpha}(r')\,,\\
    F^{(\mathrm{exc},\gamma)}_{\ell j,\tau_\alpha}\,\mathcal{R}_{\ell j}(r)
    =&\,2\,(2\gamma+1)\,\sum_{\substack{\ell_\alpha j_\alpha\\
    \nu_\alpha \nu_{\alpha'}(\in A)}} (2\ell_\alpha+1)\,(2j_\alpha+1)\,
    \varrho^{(\tau_\alpha \ell_\alpha j_\alpha)}_{\nu_\alpha \nu_{\alpha'}}\\
    &\hspace*{2cm}\times \sum_{\lambda \kappa}(2\kappa+1)\,
    \big(\ell_\alpha\,0\,\lambda\, 0\,|\,\ell\,0\big)^2
    \begin{Bmatrix} \ell_\alpha& 1/2& j_\alpha \\  \lambda& \gamma& \kappa\\
      \ell& 1/2& j\end{Bmatrix}^2 \\
    &\hspace*{4cm}\times \int r^{\prime\,2}dr'\,g_{\lambda}(r,r')\,
    R_{\nu_{\alpha'}\ell_\alpha j_\alpha}^\ast(r')\,
    R_{\nu_\alpha l_\alpha j_\alpha}(r)\,\mathcal{R}_{\ell j}(r')\,.
\end{split}\label{eq:F_C}\end{equation}

\subsection{Terms from tensor channels\label{subapp:tensor}}

The contribution of $v^{(\mathrm{TN})}$ to the SFP is
\begin{equation}\begin{split}
    U^{(\mathrm{TN,dir})}_{\ell j} &= 0\,,\\
    U^{(\mathrm{TN,exc})}_{\ell j}
    &= - \sum_{\tau_\alpha} \big\{\bar{t}^{(\mathrm{TNx})}
    +\delta_{\tau\tau_\alpha}\bar{t}^{(\mathrm{TNi})}\big\}\,
    F^{(\mathrm{exc,TN})}_{\ell j,\tau_\alpha}\,;\\
  &\quad F^{(\mathrm{exc,TN})}_{\ell j,\tau_\alpha}\,\big\{\mathcal{R}_{\ell j}(r)\,
        [Y^{(\ell)}(\hat{\mathbf{r}})\,\chi_\sigma^{(1/2)}]^{(j)}_m\big\} \\
  &\qquad = \sum_{\substack{\ell_\alpha j_\alpha m_\alpha\\
        \nu_\alpha \nu_{\alpha'}(\in A)}}
  \varrho^{(\tau_\alpha \ell_\alpha j_\alpha)}_{\nu_\alpha \nu_{\alpha'}}
  \sum_{\sigma'\tau'} \int d^3r'\,\big\{R_{\nu_{\alpha'}\ell_\alpha j_\alpha}(r')\,
  [Y^{(\ell_\alpha)}(\hat{\mathbf{r}}')\,\chi_{\sigma'}^{(1/2)}]^{(j_\alpha)}_{m_\alpha}
   \big\}^\ast\, f^{(\mathrm{TN})}\big(|\mathbf{r}-\mathbf{r}'|\big)\\
   &\hspace*{1.5cm} \times \Big\{16\pi \sum_{\lambda_1 \lambda_2}
    \delta_{\lambda_1+\lambda_2,2}\,\zeta_{\lambda_1}\,
    r^{\lambda_1} r^{\prime,\lambda_2}\,\big[Y^{(\lambda_1)}(\hat{\mathbf{r}})\,
      Y^{(\lambda_2)}(\hat{\mathbf{r}}')\big]^{(2)}\cdot
    [s^{(1)} s^{\prime(1)}]^{(2)}\Big\} \\
   &\hspace*{2cm} \times \big\{R_{\nu_\alpha\ell_\alpha j_\alpha}(r)\,
   [Y^{(\ell_\alpha)}(\hat{\mathbf{r}})\,\chi_{\sigma}^{(1/2)}]^{(j_\alpha)}_{m_\alpha}
   \big\}\,\big\{\mathcal{R}_{\ell j}(r')\,
        [Y^{(\ell)}(\hat{\mathbf{r}}')\,\chi_{\sigma'}^{(1/2)}]^{(j)}_m\big\}\,.\\
\end{split}\label{eq:U_TN}\end{equation}
The direct term vanishes because of the time-reversal symmetry
in the target w.f.
After dropping the angular-spin part,
an algebra similar to Eq.~\eqref{eq:F_exc} derives
\begin{equation}\begin{split}
    & F^{(\mathrm{exc,TN})}_{\ell j,\tau_\alpha}\,\mathcal{R}_{\ell j}(r) \\
    &\quad=\,30 \sum_{\lambda_1 \lambda_2}
    \delta_{\lambda_1+\lambda_2,2}\,(-)^{\lambda_1 +1}\,\zeta_{\lambda_1}\,
    \sqrt{(2\lambda_1+1)\,(2\lambda_2+1)}
    \sum_{\substack{\ell_\alpha j_\alpha\\
    \nu_\alpha \nu_{\alpha'}(\in A)}} (2\ell_\alpha+1)\,(2j_\alpha+1)\,
    \varrho^{(\tau_\alpha \ell_\alpha j_\alpha)}_{\nu_\alpha \nu_{\alpha'}} \\
    &\hspace*{1cm}\times \sum_{\lambda \kappa \kappa_1 \kappa_2}
    (2\kappa+1)\sqrt{(2\kappa_1+1)\,(2\kappa_2+1)}\,
    \big(\lambda\, 0\, \lambda_1\, 0\,|\, \kappa_1\, 0\big)\,
    \big(\lambda\, 0\, \lambda_2\, 0\,|\, \kappa_2\, 0\big)\\
    &\hspace*{2cm}\times
    \big(\ell_\alpha\, 0\, \kappa_1\, 0\,|\, \ell\, 0\big)\,
    \big(\ell_\alpha\, 0\, \kappa_2\, 0\,|\, \ell\, 0\big)\,
    W(2\, \lambda_2\, \kappa_1\, \lambda\,;\,
    \lambda_1\, \kappa_2)\,
    W(2\, 1\, \kappa_1\, \kappa\,;\, 1\, \kappa_2)\\
    &\hspace*{8cm}\times
    \begin{Bmatrix} \ell_\alpha & 1/2 & j_\alpha \\ \kappa_1 & 1 & \kappa \\
      \ell & 1/2 & j \end{Bmatrix}\,
    \begin{Bmatrix} \ell_\alpha & 1/2 & j_\alpha \\ \kappa_2 & 1 & \kappa \\
      \ell & 1/2 & j \end{Bmatrix}\\
    &\hspace*{5cm}\times \int r^{\prime\,2}dr'\,r^{\lambda_1} r^{\prime\,\lambda_2}\,
    g_{\lambda}(r,r')\,R_{\nu_{\alpha'}\ell_\alpha j_\alpha}^\ast(r')\,
    R_{\nu_\alpha\ell_\alpha j_\alpha}(r)\,\mathcal{R}_{\ell j}(r')\,.\\
\end{split}\label{eq:F_TN}\end{equation}

\subsection{Terms from LS channels\label{subapp:LS}}

The contribution of $v^{(\mathrm{LS})}$ to the SFP is,
after taking into account the time-reversality for the direct term,
\begin{equation}\begin{split}
  U^{(\mathrm{LS,dir})}_{\ell j} &= \sum_{\tau_\alpha} \big\{\bar{t}^{(\mathrm{LSi})}
    +\delta_{\tau\tau_\alpha}\bar{t}^{(\mathrm{LSx})}\big\}\,
    F^{(\mathrm{dir,LS})}_{\ell j,\tau_\alpha}\,;\\
    &\quad F^{(\mathrm{dir,LS})}_{\ell j,\tau_\alpha}\,\big\{\mathcal{R}_{\ell j}(r)\,
        [Y^{(\ell)}(\hat{\mathbf{r}})\,\chi_\sigma^{(1/2)}]^{(j)}_m\big\} \\
  &\qquad = \sum_{\substack{\ell_\alpha j_\alpha m_\alpha\\
        \nu_\alpha \nu_{\alpha'}(\in A)}}
  \varrho^{(\tau_\alpha \ell_\alpha j_\alpha)}_{\nu_\alpha \nu_{\alpha'}}
  \sum_{\sigma'\tau'} \int d^3r'\,\big\{R_{\nu_{\alpha'}\ell_\alpha j_\alpha}(r')\,
  [Y^{(\ell_\alpha)}(\hat{\mathbf{r}}')\,\chi_{\sigma'}^{(1/2)}]^{(j_\alpha)}_{m_\alpha}
   \big\}^\ast\\
   &\hspace*{1.5cm} \times f^{(\mathrm{LS})}\big(|\mathbf{r}-\mathbf{r}'|\big)\,
   \Big[\frac{1}{2}\big\{\boldsymbol{\ell}\cdot\mathbf{s}
    + \boldsymbol{\ell}'\cdot\mathbf{s}'
    - (\mathbf{p}\times\mathbf{s})\cdot\mathbf{r}'
    - \mathbf{r}\cdot(\mathbf{p}'\times\mathbf{s}')\big\}\Big] \\
   &\hspace*{2cm} \times \big\{R_{\nu_\alpha\ell_\alpha j_\alpha}(r')\,
   [Y^{(\ell_\alpha)}(\hat{\mathbf{r}}')\,\chi_{\sigma'}^{(1/2)}]^{(j_\alpha)}_{m_\alpha}
   \big\}\,\big\{\mathcal{R}_{\ell j}(r)\,
   [Y^{(\ell)}(\hat{\mathbf{r}})\,\chi_{\sigma}^{(1/2)}]^{(j)}_m\big\}\,,\\
  U^{(\mathrm{LS,exc})}_{\ell j} &= - \sum_{\tau_\alpha} \big\{\bar{t}^{(\mathrm{LSx})}
    +\delta_{\tau\tau_\alpha}\bar{t}^{(\mathrm{LSi})}\big\}\,
    F^{(\mathrm{exc,LS})}_{\ell j,\tau_\alpha}\,;\\
  &\quad F^{(\mathrm{exc,LS})}_{\ell j,\tau_\alpha}\,\big\{\mathcal{R}_{\ell j}(r)\,
        [Y^{(\ell)}(\hat{\mathbf{r}})\,\chi_\sigma^{(1/2)}]^{(j)}_m\big\} \\
  &\qquad = \sum_{\substack{\ell_\alpha j_\alpha m_\alpha\\
        \nu_\alpha \nu_{\alpha'}(\in A)}}
  \varrho^{(\tau_\alpha \ell_\alpha j_\alpha)}_{\nu_\alpha \nu_{\alpha'}}
  \sum_{\sigma'\tau'} \int d^3r'\,\big\{R_{\nu_{\alpha'}\ell_\alpha j_\alpha}(r')\,
  [Y^{(\ell_\alpha)}(\hat{\mathbf{r}}')\,\chi_{\sigma'}^{(1/2)}]^{(j_\alpha)}_{m_\alpha}
   \big\}^\ast\\
   &\hspace*{1.5cm} \times f^{(\mathrm{LS})}\big(|\mathbf{r}-\mathbf{r}'|\big)\,
   \Big[\frac{1}{2}\big\{\boldsymbol{\ell}\cdot\mathbf{s}
    + \boldsymbol{\ell}\cdot\mathbf{s}'
    + \boldsymbol{\ell}'\cdot\mathbf{s}
    + \boldsymbol{\ell}'\cdot\mathbf{s}'
    + (\mathbf{r}\times\mathbf{s})\cdot\mathbf{p}'
    - (\mathbf{p}\times\mathbf{s})\cdot\mathbf{r}' \\
    &\hspace*{5cm} - \mathbf{r}\cdot(\mathbf{p}'\times\mathbf{s}')
    + \mathbf{p}\cdot(\mathbf{r}'\times\mathbf{s}')\big\}\Big] \\
   &\hspace*{2cm} \times \big\{R_{\nu_\alpha\ell_\alpha j_\alpha}(r)\,
   [Y^{(\ell_\alpha)}(\hat{\mathbf{r}})\,\chi_{\sigma}^{(1/2)}]^{(j_\alpha)}_{m_\alpha}
   \big\}\,\big\{\mathcal{R}_{\ell j}(r')\,
        [Y^{(\ell)}(\hat{\mathbf{r}}')\,\chi_{\sigma'}^{(1/2)}]^{(j)}_m\big\}\,.\\
\end{split}\label{eq:U_LS}\end{equation}
The contributions of the $\boldsymbol{\ell}\cdot\mathbf{s}$
and $\boldsymbol{\ell}'\cdot\mathbf{s}'$ terms are similar
to the central channel,
because these operators only yield the constants
when acting on the w.f.'s of Eq.~\eqref{eq:basis},
\begin{equation}
  \boldsymbol{\ell}\cdot\mathbf{s} \rightarrow
  \frac{1}{2}\Big\{j(j+1) - \ell(\ell+1) - \frac{3}{4}\Big\}\,.
\label{eq:ls}\end{equation}
Owing to the symmetry in the target w.f.,
the rest of the non-vanishing direct terms have analogous forms,
and the direct SFP is expressed as
\begin{equation}\begin{split}
    F^{(\mathrm{dir,LS})}_{\ell j,\tau_\alpha}
    =& \frac{1}{4} \sum_{\substack{\ell_\alpha j_\alpha\\
    \nu_\alpha \nu_{\alpha'}(\in A)}}
    (2j_\alpha+1)\,\varrho^{(\tau_\alpha \ell_\alpha j_\alpha)}_{\nu_\alpha \nu_{\alpha'}}
    \int r^{\prime\,2}dr'\,\bigg[
      \Big\{j(j+1) - \ell(\ell+1) - \frac{3}{4}\Big\}\,
    \Big\{g_0(r,r') - \frac{r'}{3r}g_1(r,r')\Big\}\, \\
    &\hspace*{5cm}+ \Big\{j_\alpha(j_\alpha+1)
    - \ell_\alpha(\ell_\alpha+1) - \frac{3}{4}\Big\}\,
      \Big\{g_0(r,r') - \frac{r}{3r'}g_1(r,r')\Big\}\bigg]\\
    &\hspace*{4.5cm}\times R_{\nu_{\alpha'}\ell_\alpha j_\alpha}^\ast(r')\,
    R_{\nu_\alpha \ell_\alpha j_\alpha}(r')\,,\\
\end{split}\label{eq:F_LS_dir}\end{equation}
Finally, the exchange part of the LS channel is obtained
via elaborate algebras on each term appearing in Eq.~\eqref{eq:U_LS},
\begin{equation}\begin{split}
    & F^{(\mathrm{exc,LS})}_{\ell j,\tau_\alpha}\,\mathcal{R}_{\ell j}(r) \\
    &\quad= \frac{1}{2} \sum_{\substack{\ell_\alpha j_\alpha\\
    \nu_\alpha \nu_{\alpha'}(\in A)}}(2\ell_\alpha+1)\,(2j_\alpha+1)\,
    \varrho^{(\tau_\alpha \ell_\alpha j_\alpha)}_{\nu_\alpha \nu_{\alpha'}}\,
    \sum_{\lambda}(2\lambda+1)\,
    \big(\ell_\alpha\,0\,\lambda\, 0\,|\,\ell\,0\big)^2
    \begin{Bmatrix} \ell_\alpha& 1/2& j_\alpha \\  \lambda& 0& \lambda\\
      \ell& 1/2& j\end{Bmatrix}^2 \\
    &\hspace*{1cm}\times \Big\{j(j+1) - \ell(\ell+1) - \frac{3}{4}
    + j_\alpha(j_\alpha+1)- \ell_\alpha(\ell_\alpha+1) - \frac{3}{4}\Big\}\,\\
    &\hspace*{4cm}\times \int r^{\prime\, 2}dr'\,g_{\lambda}(r,r')\,
    R_{\nu_{\alpha'}\ell_\alpha j_\alpha}^\ast(r')\,
    R_{\nu_\alpha l_\alpha j_\alpha}(r)\,\mathcal{R}_{\ell j}(r')\,\\
    &\quad+ \sum_{\substack{\ell_\alpha j_\alpha\\
    \nu_\alpha \nu_{\alpha'}(\in A)}} (2\ell_\alpha+1)\,(2j_\alpha+1)\,
    \varrho^{(\tau_\alpha \ell_\alpha j_\alpha)}_{\nu_\alpha \nu_{\alpha'}}\,
    \sum_{\lambda\lambda'\lambda''\lambda'''\kappa} (2\kappa+1)\,
    \big(\ell_\alpha\, 0\, \lambda''\, 0\,|\ell\, 0\big)\,
    \big(\ell_\alpha\, 0\, \lambda'''\, 0\,|\ell\, 0\big)\,\\
    &\hspace*{1cm}\times
    \begin{Bmatrix} \ell_\alpha& 1/2& j_\alpha\\ \kappa& 0& \kappa\\
      \ell& 1/2& j\end{Bmatrix}\,
    \begin{Bmatrix} \ell_\alpha& 1/2& j_\alpha\\ \lambda'& 1& \kappa\\
      \ell& 1/2& j\end{Bmatrix}\,
      \int r^{\prime\, 2}dr'\,g_\lambda(r,r')\,
      R_{\nu_{\alpha'}\ell_\alpha j_\alpha}^\ast(r') \\
    &\hspace*{1.5cm}\times \bigg[
     \frac{\sqrt{3}}{2} \sqrt{\ell\,(\ell+1)\,(2\ell+1)}\,
     \bigg(\sqrt{2\kappa+1}\,\delta_{\lambda\lambda'}\delta_{\lambda\lambda''}
     \delta_{\lambda\lambda'''}\,
     W(\ell\,1\,\ell_\alpha\,\lambda\,;\,\ell\,\kappa) \\
     &\hspace*{2.5cm}- \frac{6r}{r'}\sqrt{(2\lambda'+1)\,(2\lambda''+1)}\,
    \big(\lambda\, 0\, 1\, 0\,|\,\lambda''\, 0\big)\,
    \big(\lambda\, 0\, 1\, 0\,|\,\lambda'''\, 0\big)\,
    W(\lambda\, 1\, \kappa\, 1\,;\, \lambda'\, 1)\\
    &\hspace*{3cm}\times \Big\{ \sqrt{2\kappa+1}\,\delta_{\lambda'\lambda'''}\,
    W(\ell\,1\,\ell_\alpha\,\lambda''\,;\,\ell\,\kappa)\,
    W(\lambda\, 1\, \kappa\, 1\,;\, \lambda''\, 1)\\
    &\hspace*{3.5cm}+ \sqrt{2\lambda'+1}\,\delta_{\kappa\lambda'''}\,
    W(\ell\,1\,\ell_\alpha\,\lambda''\,;\,\ell\,\lambda')\,
    W(\lambda\, 1\, \lambda'\, 1\,;\, \lambda''\, 1) \Big\}\bigg)\,
      R_{\nu_\alpha\ell_\alpha j_\alpha}(r)\,\mathcal{R}_{\ell j}(r') \\
    &\hspace*{2cm}- (-)^{\lambda'+\kappa}
    \frac{\sqrt{3}}{2} \sqrt{l_\alpha\,(l_\alpha+1)\,(2l_\alpha+1)}\,
     \bigg(\sqrt{2\kappa+1}\,\delta_{\lambda\lambda'}\delta_{\lambda\lambda''}
     \delta_{\lambda\lambda'''}\,
     W(\ell_\alpha\,1\,\ell\,\lambda\,;\,\ell_\alpha\,\kappa) \\
     &\hspace*{2.5cm}- \frac{6r'}{r}\sqrt{(2\lambda'+1)\,(2\lambda''+1)}\,
    \big(\lambda\, 0\, 1\, 0\,|\,\lambda''\, 0\big)\,
    \big(\lambda\, 0\, 1\, 0\,|\,\lambda'''\, 0\big)\,
    W(\lambda\, 1\, \kappa\, 1\,;\, \lambda'\, 1)\\
    &\hspace*{3cm}\times \Big\{ \sqrt{2\kappa+1}\,\delta_{\lambda'\lambda'''}\,
    W(\ell_\alpha\,1\,\ell\,\lambda''\,;\,\ell_\alpha\,\kappa)\,
    W(\lambda\, 1\, \kappa\, 1\,;\, \lambda''\, 1)\\
    &\hspace*{3.5cm}+ \sqrt{2\lambda'+1}\,\delta_{\kappa\lambda'''}\,
    W(\ell_\alpha\,1\,\ell\,\lambda''\,;\,\ell_\alpha\,\lambda')\,
    W(\lambda\, 1\, \lambda'\, 1\,;\, \lambda''\, 1) \Big\}\bigg)\,
      R_{\nu_\alpha\ell_\alpha j_\alpha}(r)\,\mathcal{R}_{\ell j}(r') \\
    &\hspace*{2cm}-3\sqrt{2\,(2\lambda'+1)}\,
      \delta_{\kappa\lambda''} \delta_{\lambda'\lambda'''}\,
    \big(\lambda\, 0\, 1\, 0\,|\,\lambda''\, 0\big)\,
    \big(\lambda\, 0\, 1\, 0\,|\,\lambda'''\, 0\big) \\
    &\hspace*{3cm}\times W(\lambda\, 1\, \kappa\, 1\,;\, \lambda'\, 1)\,
    \Big\{r\,R_{\nu_\alpha\ell_\alpha j_\alpha}(r)\,\frac{d\mathcal{R}_{\ell j}(r')}{dr'}
    - (-)^{\lambda'+\kappa}\,r'\,\frac{dR_{\nu_\alpha\ell_\alpha j_\alpha}(r)}{dr}\,
    \mathcal{R}_{\ell j}(r')\Big\}\bigg] \,.
\end{split}\label{eq:F_LS_exc}\end{equation}
The derivative in the term including $d\mathcal{R}_{\ell j}/dr'$
can be transferred to the derivative of $g_\lambda$
and $R_{\nu_{\alpha'}\ell_\alpha j_\alpha}^\ast$ via integration by parts.

\subsection{Terms from density-dependent channels\label{subapp:DD}}

Because the $v_{\alpha\beta}^{(\mathrm{C}\rho)}$
and $v_{\alpha\beta}^{(\mathrm{LS}\rho)}$ terms in Eq.~\eqref{eq:effint}
contain the delta function $\delta(\mathbf{r}_{\alpha\beta})$,
their contributions to the SFP are local.
They resemble the forms known in the Skyrme HF potential~\cite{ref:CNP1,
  ref:Nak15},
\begin{equation}\begin{split}
    U^{(\mathrm{C}\rho)} =& \frac{1}{4} \bigg[
      \Big\{C_\mathrm{SE}[\rho(r)] + 3\,C_\mathrm{TE}[\rho(r)]\Big\}\,\rho(r)
      + \Big\{C_\mathrm{SE}[\rho(r)] - 3\,C_\mathrm{TE}[\rho(r)]\Big\}\,
      \rho_\tau(r) \bigg]\\
    & + \frac{1}{8} \bigg[
      \Big\{\frac{\partial C_\mathrm{SE}[\rho(r)]}{\partial\rho}
      + 3\,\frac{\partial C_\mathrm{TE}[\rho(r)]}{\partial\rho}\Big\}\,
      \big\{\rho(r)\big\}^2
      + \Big\{\frac{\partial C_\mathrm{SE}[\rho(r)]}{\partial\rho}
      - 3\,\frac{\partial C_\mathrm{TE}[\rho(r)]}{\partial\rho}\Big\}
      \sum_{\tau_\alpha}\big\{\rho_{\tau_\alpha}(r)\big\}^2 \bigg]\,,\\
    U^{(\mathrm{LS}\rho)}_{\ell j} =& 
    - \frac{1}{2}\,D[\rho(r)]\Big\{
      \Big(\frac{d}{dr}+\frac{2}{r}\Big)\mathcal{J}(r)\,
      + \Big(\frac{d}{dr}+\frac{2}{r}\Big)\mathcal{J}_\tau(r)\Big\} \\
    & - \frac{1}{4}\,\frac{\partial D[\rho(r)]}{\partial\rho}\,\bigg[
      \rho(r)\,\Big(\frac{d}{dr}+\frac{2}{r}\Big)\mathcal{J}(r)
      + \sum_{\tau_\alpha} \rho_{\tau_\alpha}(r)\,
      \Big(\frac{d}{dr}+\frac{2}{r}\Big)\mathcal{J}_{\tau_\alpha}(r) \\
    &\hspace*{3cm}+ \mathcal{J}_\tau(r)\,\frac{d}{dr}\rho(r)
      - \sum_{\tau_\alpha} \mathcal{J}_{\tau_\alpha}(r)\,
      \frac{d}{dr}\rho_{\tau_\alpha}(r) \bigg] \\
    & + \frac{1}{2r}\,\bigg[ D[\rho(r)]\,
      \frac{d}{dr}\big\{\rho(r)+\rho_\tau(r)\big\}
      + \frac{1}{2}\,\frac{\partial D[\rho(r)]}{\partial\rho}\,
      \big\{\rho(r)+\rho_\tau(r)\big\}\,\frac{d}{dr}\rho(r)\bigg]\\
    &\hspace*{8cm}\times \Big\{j(j+1)- \ell(\ell+1) - \frac{3}{4}\Big\}\,,
\end{split}\end{equation}
where
\begin{equation}\begin{split}
      \rho_\tau(r)
      &= \frac{1}{4\pi} \sum_{\substack{\ell_\alpha j_\alpha\\
          \nu_\alpha \nu_{\alpha'}(\in A)}}
      (2j_\alpha+1)\,\varrho^{(\tau \ell_\alpha j_\alpha)}_{\nu_\alpha \nu_{\alpha'}}\,
      R_{\nu_{\alpha'}\ell_\alpha j_\alpha}^\ast(r)\,R_{\nu_\alpha\ell_\alpha j_\alpha}(r)\,,\quad
      \rho(r) = \sum_\tau \rho_\tau(r)\,,\\
      \mathcal{J}_\tau(r)
      &= \frac{1}{4\pi} \sum_{\substack{\ell_\alpha j_\alpha\\
          \nu_\alpha \nu_{\alpha'}(\in A)}}
      (2j_\alpha+1)\,
      \Big\{j_\alpha(j_\alpha+1)- \ell_\alpha(\ell_\alpha+1) - \frac{3}{4}\Big\}\,
      \varrho^{(\tau \ell_\alpha j_\alpha)}_{\nu_\alpha \nu_{\alpha'}}\,\frac{1}{r}\,
      R_{\nu_{\alpha'}\ell_\alpha j_\alpha}^\ast(r)\,R_{\nu_\alpha\ell_\alpha j_\alpha}(r)\,,\\
      &\hspace*{9cm} \mathcal{J}(r) = \sum_\tau \mathcal{J}_\tau(r)\,.\\
\end{split}\end{equation}

\subsection{Forms of $g_\lambda$\label{subapp:f_lambda}}

We here present the forms of $g_\lambda$ of Eq.~\eqref{eq:f_lambda}
for the Gauss and Yukawa functions.
The Fourier transform helps derive $g_\lambda$.
Because
\begin{equation}\begin{split}
  f(r_{\alpha\beta}) &= \frac{1}{(2\pi)^3} \int d^3q\,
  \tilde{f}(q)\,e^{i\mathbf{q}\cdot\mathbf{r}_{\alpha\beta}} \\
  &= \frac{2}{\pi} \sum_\lambda \int_0^\infty q^2 dq\,
  \tilde{f}(q)\,j_\lambda(qr_\alpha)\,j_\lambda(qr_\beta)\,
  Y^{(\lambda)}(\hat{\mathbf{r}}_\alpha)\cdot
  Y^{(\lambda)}(\hat{\mathbf{r}}_\beta)\,,
\end{split}\end{equation}
where
\begin{equation}
  \tilde{f}(q) = \int d^3r\,f(r)\,e^{-i\mathbf{q}\cdot\mathbf{r}}\,,
\end{equation}
$g_\lambda$ can be calculated as
\begin{equation}
  g_\lambda(r_\alpha,r_\beta)
  = \frac{2\lambda+1}{2\pi^2} \int_0^\infty q^2 dq\,
  \tilde{f}(q)\,j_\lambda(qr_\alpha)\,j_\lambda(qr_\beta)\,.
\label{eq:g_lambda}\end{equation}
We obtain, for the Gauss function $f(r_{\alpha\beta})=e^{-(\mu r_{\alpha\beta})^2}$,
\begin{equation}
  g_\lambda(r_\alpha,r_\beta)
  = \frac{\sqrt{\pi}(2\lambda+1)}{2\mu\sqrt{r_\alpha r_\beta}}\,
  e^{-\mu^2(r_\alpha^2+r_\beta^2)}\,I_{\lambda+1/2}(2\mu^2 r_\alpha r_\beta)\,,
\end{equation}
and, for the Yukawa function $f(r_{\alpha\beta})=e^{-\mu r_{\alpha\beta}}
/\mu r_{\alpha\beta}$,
\begin{equation}
  g_\lambda(r_\alpha,r_\beta)
  = \frac{2\lambda+1}{\mu\sqrt{r_\alpha r_\beta}}\,
  I_{\lambda+1/2}(\mu r_<)\,K_{\lambda+1/2}(\mu r_>)\,;
  \quad r_< = \min(r_\alpha,r_\beta)\,,~r_> = \max(r_\alpha,r_\beta)\,.
\end{equation}
Here $I_\nu(z)$ and $K_\nu(z)$ are the modified Bessel functions.

For $f(r_{\alpha\beta})=1/r_{\alpha\beta}$ that appears in the Coulomb interaction,
the following well-known result is obtained
from Eq.~\eqref{eq:g_lambda},
\begin{equation}
  g_\lambda(r_\alpha,r_\beta) = \frac{r_<^\lambda}{r_>^{\lambda+1}}\,.
\end{equation}

\section{Center-of-mass correction\label{app:cm}}

We here discuss the influence of $H_\mathrm{c.m.}$ in Eq.~\eqref{eq:Hamil}.
Let $A$ be the mass number of the target nucleus and $A'=A+1$.
We denote the Hamiltonian and the momentum of the target nucleus
by $H_A$ and $\mathbf{P}_A$.
The momentum of the scattered nucleon relative to the target $A$ is defined by
\begin{equation}
\tilde{\mathbf{p}}_N := \frac{1}{A+1}\big(A\,\mathbf{p}_N - \mathbf{P}_A\big)\,,
\label{eq:relmom}\end{equation}
yielding
\begin{equation}
  \mathbf{p}_N = \Big(1+\frac{1}{A}\Big)\tilde{\mathbf{p}}_N
  + \frac{1}{A}\,\mathbf{P}_A\,,\quad
  \mathbf{P} = \Big(1+\frac{1}{A}\Big)
  \big(\mathbf{P}_A+\tilde{\mathbf{p}}_N\big)\,.
\label{eq:totmom}\end{equation}
The center-of-mass (c.m.) Hamiltonian in Eq.~\eqref{eq:Hamil}
is then rewritten as
\begin{equation}
H_\mathrm{c.m.} = \frac{A+1}{2A^2M}\big(\mathbf{P}_A+\tilde{\mathbf{p}}_N\big)^2\,,
\end{equation}
and we obtain
\begin{equation}
  \frac{\mathbf{p}_N^2}{2M} - H_\mathrm{c.m.}
  = \frac{1}{2M}\Big(1+\frac{1}{A}\Big)\tilde{\mathbf{p}}_N^2
  - \frac{\mathbf{P}_A^2}{2AM}\,.
\label{eq:cmcorr}\end{equation}
By including the second term on the rhs of Eq.~\eqref{eq:cmcorr}
in the nuclear structure calculation with $H_A$,
the correction factor $(1+1/A)$ to the first term,
which is like the reduced mass
but does not involve the binding energy of $A$,
makes the c.m. correction to the Schr\"{o}dinger equation
for the scattering wave.

\begin{acknowledgments}

Discussions with D.T.~Khoa, M.~Kohno and K.~Sumiyoshi
are gratefully acknowledged.
A part of the numerical calculations has been performed on HITAC SR24000
at Institute of Management and Information Technologies in Chiba University.
\end{acknowledgments}


\end{document}